\documentclass[aps,nofootinbib,floatfix,showpacs,preprintnumbers,twocolumn]{revtex4} 
\usepackage{graphicx}
\usepackage{bm}
\usepackage{mathrsfs}
\usepackage{float}
\usepackage{tabularx}
\usepackage{amsmath,amsfonts,amssymb}

\input epsf

\graphicspath{{/Users/lstrigar/Dropbox/NeutrinoBackground/sterile/figures/}}

\def\lsim{\mathrel{\raise.3ex\hbox{$<$\kern-.75em\lower1ex\hbox{$\sim$}}}}
\def\gsim{\mathrel{\raise.3ex\hbox{$>$\kern-.75em\lower1ex\hbox{$\sim$}}}}

\def\cmm2{{\,\rm cm^{-2}}}
\def\cm2{{\,{\rm cm}^2}}
\def\cmm3{{\,{\rm cm}^{-3}}}
\def\gcmm3{{\,{\rm g\,cm^{-3}}}}

\def\fun#1#2{\lower3.6pt\vbox{\baselineskip0pt\lineskip.9pt
  \ialign{$\mathsurround=0pt#1\hfil##\hfil$\crcr#2\crcr\sim\crcr}}}

\def\be{\begin{equation}}
\def\ee{\end{equation}}
\def\bea{\begin{eqnarray}}
\def\eea{\end{eqnarray}}

\begin{document}

\title{Complementarity of dark matter detectors in light of the neutrino background}
\author{F. Ruppin}\email{ruppin@mit.edu} \affiliation{Department of Physics, Massachusetts Institute of Technology, Cambridge, MA 02139, USA}
\author{J.~Billard}\email{billard@mit.edu} \affiliation{Department of Physics, Massachusetts Institute of Technology, Cambridge, MA 02139, USA}
\author{L.~Strigari}  \affiliation{Department of Physics, Indiana University, Bloomington, IN 47405-7105, USA}
\author{E. Figueroa-Feliciano} \affiliation{Department of Physics, Massachusetts Institute of Technology, Cambridge, MA 02139, USA}

\smallskip
\begin{abstract}

Direct detection dark matter experiments looking for WIMP-nucleus elastic scattering will soon be sensitive to an irreducible background from neutrinos which will drastically affect their discovery potential. Here we explore how the neutrino background will affect future ton-scale experiments considering both spin-dependent and spin-independent interactions. We show that combining data from experiments using different targets can improve the dark matter discovery potential due to target complementarity. We find that in the context of spin-dependent interactions, combining results from several targets can greatly enhance the subtraction of the neutrino background for WIMP masses below 10 GeV/c$^2$ and therefore probe dark matter models to lower cross-sections. In the context of target complementarity, we also explore how one can tune the relative exposures of different target materials to optimize the WIMP discovery potential.

\end{abstract}
\maketitle

\section{Introduction}
\label{sec:intro}

Numerous independent surveys have demonstrated evidence on both cosmological and galactic scales that about 30\% of the matter energy density of the Universe consists of non-baryonic, non-luminous matter. A leading candidate for this dark matter is a yet-to-be-discovered weakly interactive massive particle (WIMP) which could directly interact with detectors based on Earth leading to keV-scale nuclear recoils. Direct dark matter detection experiments are now probing well-motivated models of extensions to the Standard Model such as Supersymmetry which naturally predict dark matter candidates~\cite{Jungman:1995df,Bertone:2004pz,Strigari:2013iaa}. 

As the exposures of direct dark matter detection experiments continue to improve, they will soon have enough sensitivity to detect neutrinos from several astrophysical sources such as the Sun, the atmosphere, and diffuse supernovae \cite{neutrino_Cabrera,neutrino_Fisher,LouisNeutrino,neutrino_Gutlein,neutrino_Harnik,Gutlein:2014gma}. For example, a 1~keV threshold Xe based experiment with a 1~ton-year exposure will detect about 100 ${}^{8}\rm{B}$ solar neutrino events via coherent neutrino-nucleus scattering (CNS). In fact for some WIMP masses, such neutrino backgrounds can almost perfectly mimic a WIMP signal. It has been shown in Ref.~\cite{Billard:2013qya} that the CNS background leads to a strong reduction of the discovery potential of upcoming experiments. Therefore, though neither coherent neutrino scattering nor the WIMP-nucleus interaction have conclusively been observed yet, the search for discrimination methods to disentangle WIMPs from neutrino events is a necessity. 

Several methods to improve on the discrimination power between a WIMP and a neutrino origin of the observed nuclear recoils have been suggested~\cite{Billard:2013qya}, including using the annual modulation signal~\cite{modulation} or directional detection methods~\cite{directionality_Ahlen,directionality_Grothaus}. In this paper, we propose a new method to reduce the effect of this neutrino background by looking for a possible complementarity between different target nuclei. In addition to discussing the target complementarity for different spin-independent (SI) targets, for the first time we discuss the prospects for complementarity using spin-dependent (SD) targets. 

This paper is organized as follows. In Sec. \ref{sec:neutfloor}, we briefly review how to compute the WIMP and neutrino event rates and we explain the test statistic that has been used to generate discovery limits. We then compute some discovery limits for several upcoming direct dark matter detection experiments in light of the neutrino background. In Sec. \ref{sec:beyond}, we show how target complementarity can help on improving the WIMP discovery potential by considering both SI and SD interactions. We then illustrate the improvement on the discovery limits by combining data from different targets. In Sec. \ref{sec:optimization}, we explore how one can optimize the relative exposures of the different target materials in a given experiment in order to maximize its discovery potential using the effect of target complementarity.

\section{The neutrino background}
\label{sec:neutfloor}

\subsection{Event rates computation}

 \begin{table*}
\begin{ruledtabular}
\begin{tabular}{|c|ccccccc|}
   & \bf{Nucleus} & \bf{mean A} & \bf{Z} & \bf{Isotopic fraction} & \bf{J} & $\mathbf{\langle S_p \rangle}$ & $\mathbf{\langle S_n \rangle}$\\
\hline
 & $\rm{W}$ & 183.91 & 74 & 1.0 & & & \\
 & $\rm{Xe}$ & 131.29 & 54 & 1.0 & & & \\
 & $\rm{I}$ & 127.00 & 53 & 1.0 & & & \\
 & $\rm{Ge}$ & 72.63 & 32 & 1.0 & & & \\
SI & $\rm{Ca}$ & 40.12 & 20 & 1.0 & & not considered & \\
 & $\rm{Ar}$ & 39.98 & 18 & 1.0 & & & \\
 & $\rm{Si}$ & 28.11 & 14 & 1.0 & & & \\
 & $\rm{F}$ & 19.00 & 9 & 1.0 & & & \\
 & $\rm{O}$ & 16.00 & 8 & 1.0 & & & \\
 & $\rm{C}$ & 12.01 & 6 & 1.0 & & & \\
 \hline
& \bf{Nucleus} & \bf{A} & \bf{Z} & \bf{Isotopic fraction} & \bf{J} & $\mathbf{\langle S_p \rangle}$ & $\mathbf{\langle S_n \rangle}$\\
 \hline
 & $\rm{Xe}$ & 131 & 54 & 0.2129 & 3/2 & -0.009 & -0.227\\
 & $\rm{Xe}$ & 129 & 54 & 0.264 & 1/2 & 0.028 & 0.359\\
SD & $\rm{I}$ & 127 & 53 & 1.0 & 5/2 & 0.309 & 0.075\\
 & $\rm{Ge}$ & 73 & 32 & 0.0776 & 9/2 & 0.030 & 0.378\\
 & $\rm{Si}$ & 29 & 14 & 0.0468 & 1/2 & -0.002 & 0.130\\
 & $\rm{F}$ & 19 & 9 & 1.0 & 1/2 & 0.477 & -0.004\\
\end{tabular}
\caption{Considered target nuclei properties used for this study for both spin-dependent (SI) and spin-independent (SD) interactions. For SI interactions, a single ``isotope'' with the mean atomic weight of the target was used. Nuclear properties are taken from~\cite{table}.\label{tab:nuclei}}
\end{ruledtabular}
\end{table*}
\begin{table}
\begin{ruledtabular}
\begin{tabular}{|cccc|}
$\mathbf{\nu}$ \bf{type} & $\mathbf{E_{\nu}^{\rm{max}}}$ \bf{(MeV)} & $\mathbf{E_{r_{\rm{Ge}}}^{\rm{max}}}$ \bf{(keV)} & $\mathbf{\nu}$ \bf{flux}\\
 & & & $\mathbf{(\rm{cm^{-2}.s^{-1}})}$\\
\hline
pp & 0.42341 & $5.30\times 10^{-3}$ & $5.99\pm 0.06\times 10^{10}$\\
${}^{7}\rm{Be}$ & 0.861 & 0.0219 & $4.84\pm 0.48\times 10^9$\\
pep & 1.440 & 0.0613 & $1.42\pm 0.04\times 10^8$\\
${}^{15}\rm{O}$ & 1.732 & 0.0887 & $2.33\pm 0.72\times 10^8$\\
${}^{8}\rm{B}$ & 16.360 & 7.91 & $5.69\pm 0.91\times 10^6$\\
hep & 18.784 & 10.42 & $7.93\pm 1.27\times 10^3$\\
DSNB & 91.201 & 245 & $85.5\pm 42.7$\\
Atm. & 981.748 & $27.7\times 10^3$ & $10.5\pm 2.1$\\
\end{tabular}
\caption{Relevant neutrino fluxes to the background of direct dark matter detection experiments. Also shown are the respective maximum neutrino energy, maximum recoil energy on a Ge target, and overall fluxes and uncertainties \cite{Bahcall:2004pz,atmosneutrino,DSNBneutrino}.\label{tab:neutrino}}
\end{ruledtabular} 
\end{table}

In this section, we review how to compute the WIMP event rates for both SI and SD interactions and the coherent neutrino event rates induced by the solar, atmospheric and diffuse supernovae neutrinos \cite{neutrino_Cabrera,neutrino_Fisher,LouisNeutrino,neutrino_Gutlein,neutrino_Harnik,Gutlein:2014gma}. 

Direct dark matter detection aim to detect elastic scattering between a WIMP from the galactic halo and the detector material. In the non-relativistic limit, the interaction between a WIMP and a nucleus is well described by the superposition of a spin-dependent (SD) and a spin-independent (SI) contribution to the total cross section. When the transferred energy is equal to 0 the WIMP-nucleus cross section ($\sigma_0^{SI,SD}$) is related to the WIMP-nucleon normalized cross section as \cite{formfactor}
\begin{equation}
\sigma_{SI}^{p,n} = \frac{\mu_p^2}{\mu_N^2} \times \frac{1}{A^2} \times \sigma_0^{SI}({}^{A}X)
\label{eq:sigSI}
\end{equation} 
\begin{equation}
\sigma_{SD}^{p,n} = \frac{3}{4} \times \frac{\mu_p^2}{\mu_N^2} \times \frac{J}{J+1} \times \frac{1}{\langle S_{p,n} \rangle^2}\sigma_0^{SD}({}^{A}X)
\label{eq:sigSD}
\end{equation}
where ${\mu_N = m_{\chi}m_N/(m_{\chi}+m_N)}$ is the WIMP-nucleus reduced mass, $A$ is the number of nucleons in the considered target nucleus and $\mu_p$ is the WIMP-nucleon reduced mass. For the spin-dependent cross section (eq.~\ref{eq:sigSD}), $J$ corresponds to the total angular momentum of the nucleus and $\langle S_{p,n} \rangle$ are its mean spin content. 

\begin{figure*}
\begin{center}
\includegraphics[width=0.95\columnwidth]{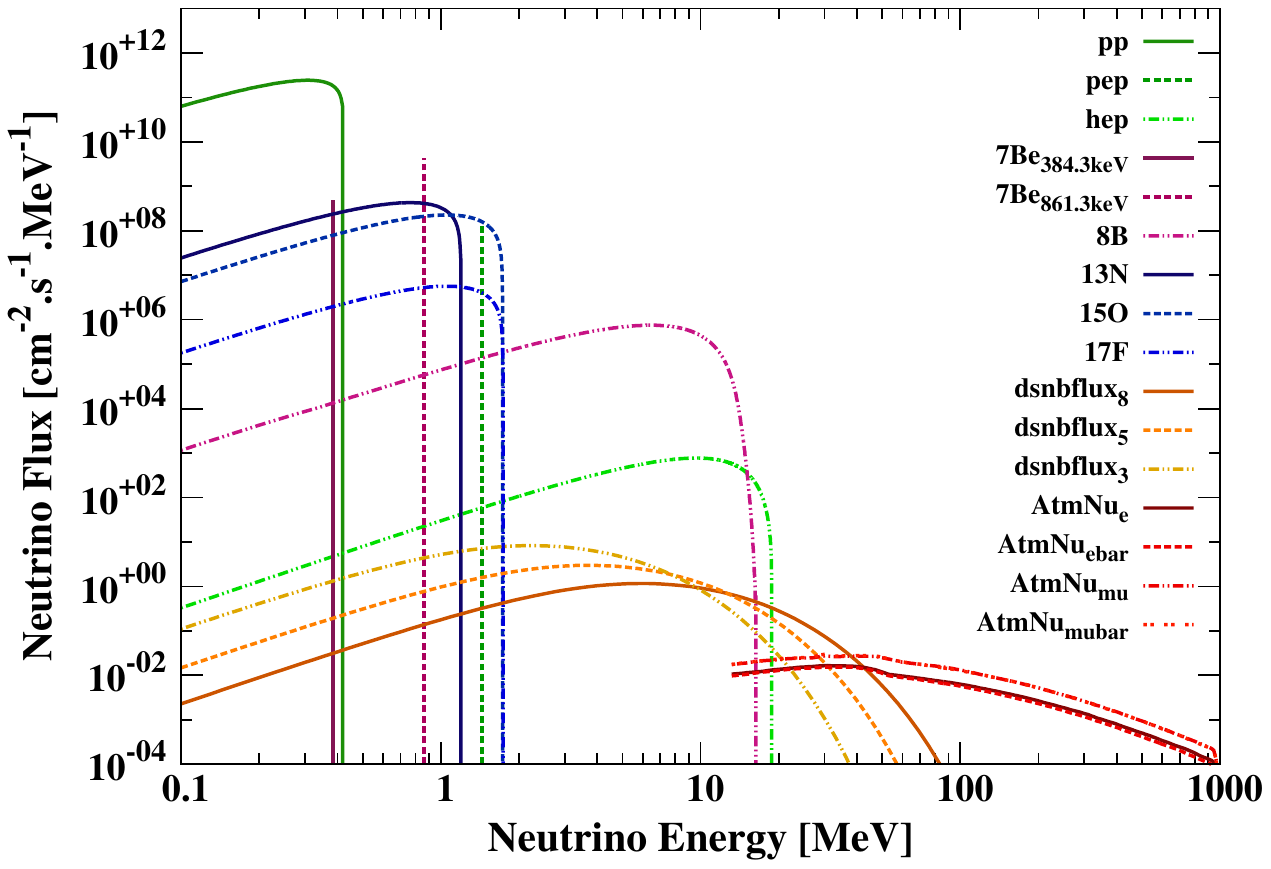}
\hspace{0.5cm}
\includegraphics[width=0.95\columnwidth]{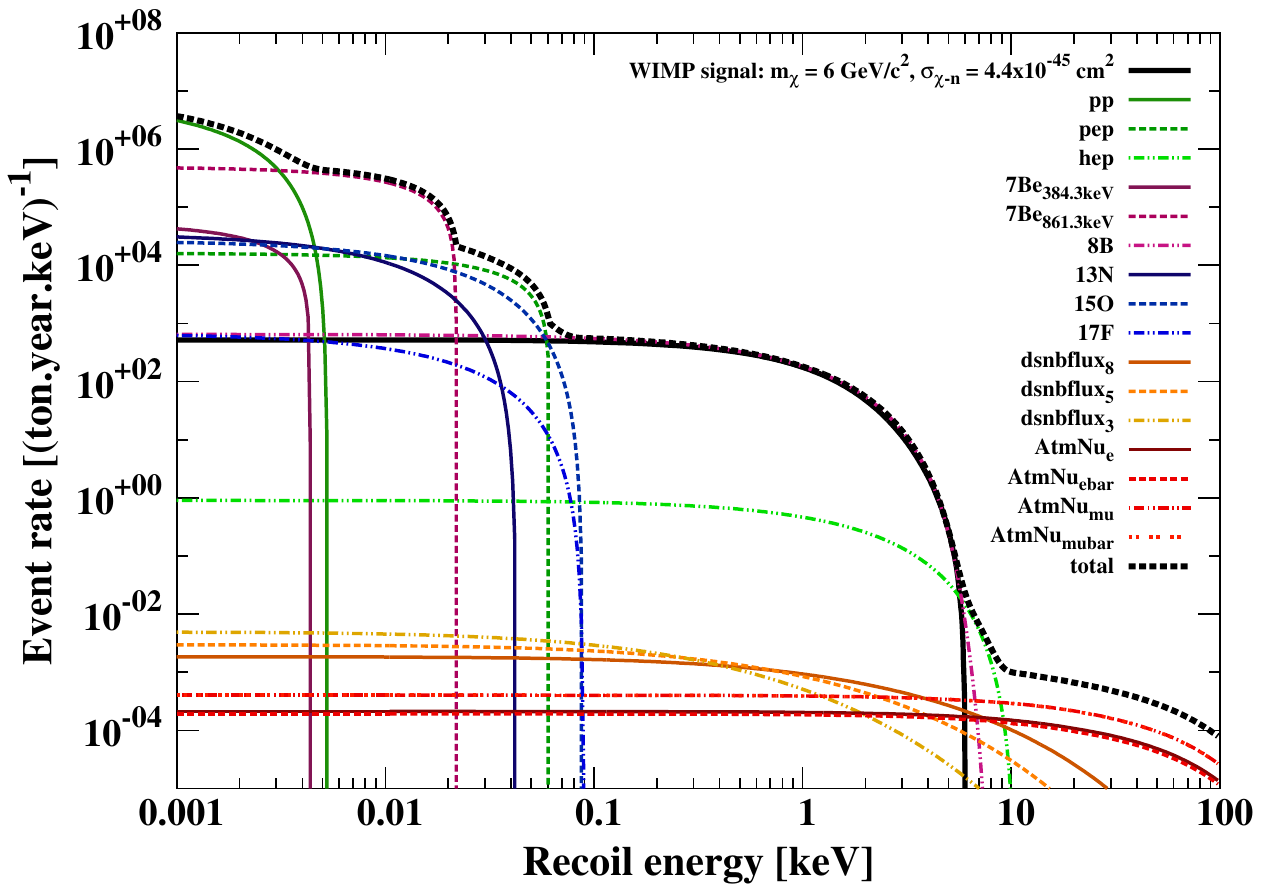}
\end{center}
\caption{Left: Relevant neutrino fluxes to the background of direct dark matter detection experiments: Solar, atmospheric, and diffuse supernovae \cite{Bahcall:2004pz,atmosneutrino,DSNBneutrino}. Right: Neutrino background event rates for a germanium based detector. The black dashed line corresponds to the sum of the neutrino induced nuclear recoil event rates. Also shown is the similarity between the event rate from a $6~\rm{GeV/c^2}$ WIMP with a SI cross section on the nucleon of $4.4\times 10^{-45}~\rm{cm^2}$ (black solid line) and the ${}^{8}\rm{B}$ neutrino event rate.}
\label{fig:eventrate}
\end{figure*}

In this study, we only consider the simplest cases in which the WIMP-nucleon cross section is due either solely to a SI interaction or to a SD interaction. More generally, the interaction of a nucleus which has a non zero total angular momentum with a WIMP is a superposition of these SI and SD terms. Isotopes of a given target which have a zero total angular momentum cannot be involved in a SD interaction. Table~\ref{tab:nuclei} shows the different target properties that we use~\cite{table}. Note that the mean spin content values depend on the considered nuclear model. The target properties used in this study are similar to within a few percent from those found in Ref.~\cite{spincontent}. The WIMP event rate as a function of the recoil energy of a given target element ${}^{A}_{Z}\rm{X}$ is given by:
\begin{equation}
\frac{dR_{\chi}}{dE_r} = MT \sum_{A} f_{A} \frac{\rho_0\sigma_0^{SI,SD}({}^{A}X)}{2m_{\chi}\mu_N^2}F_{SI,SD}^2(E_r) \int_{v_{min}}\frac{f(\vec{v})}{v}d^3\vec{v}
\end{equation}  
where $M$ is the detector mass, $T$ is the time of exposition, $f_{A}$ is the isotopic fraction of ${}^{A}_{Z}\rm{X}$ (see Table~\ref{tab:nuclei}), $\rho_0 = 0.3~\rm{GeV/c^2/cm^3}$ is the standard value for the local density of dark matter, and $v_{min}$ is the minimum velocity required for a WIMP to induce a nuclear recoil with a  recoil energy $E_r$. For the sake of comparison with existing WIMP constraints, we consider the standard Maxwell-Boltzmann WIMP velocity distribution characterized by a dispersion $\sigma_v$ related to the local circular velocity $v_0$ such that $\sigma_v = v_0/\sqrt{2}$ and an escape velocity $v_{\rm esc}$ = 544 km/s. $F_{SI,SD}(E_r)$ are the SI and SD form factors that describe the coherence of the WIMP-nucleus interaction. In the following, we will consider the standard Helm form factor for the SI interaction and the thin shell approximation for the SD interaction form factor from Ref.~\cite{formfactor}.\\

\indent Similar to the WIMP event rate calculation, the neutrino event rate is computed by the convolution of the neutrino-nucleus cross section with the neutrino flux as
\begin{equation}
\frac{dR_{\nu}}{dE_r} = MT \times \sum_{A} f_{A} \int_{E_\nu^{\rm min}} \frac{dN}{dE_{\nu}} \frac{d\sigma(E_{\nu},E_r)}{dE_r}dE_{\nu}
\end{equation}
where $\frac{dN}{dE_{\nu}}$ corresponds to the neutrino flux. As it has been shown in Ref.~\cite{neutrinoAsquared}, the neutrino-nucleon elastic interaction is theoretically well-understood within the Standard Model, and leads to a coherence effect implying a neutrino-nucleus cross section that approximately scales as the atomic number ($A$) squared when the momentum transfer is below a few keV. At tree level, the neutrino-nucleon elastic scattering is a neutral current interaction that proceeds via the exchange of a $Z$ boson. The resulting differential neutrino-nucleus cross section as a function of the recoil energy and the neutrino energy is given by \cite{neut_nucleus_cs}:
\begin{equation}
\frac{d\sigma(E_{\nu},E_r)}{dE_r} = \frac{G_f^2}{4\pi}Q_{\omega}^2m_N\left(1-\frac{m_NE_r}{2E_{\nu}^2}\right)F^2_{SI}(E_r)
\end{equation}
where $m_N$ is the nucleus mass, $G_f$ is the Fermi coupling constant and $Q_{\omega} = N - (1-4\sin^2\theta_{\omega})Z$ is the weak nuclear hypercharge with $N$ the number of neutrons, $Z$ the number of protons, and $\theta_{\omega}$ the weak mixing angle. The presence of the form factors describes the loss of coherence at higher momentum transfer and is assumed to be the same as for the WIMP-nucleus SI scattering. Interestingly, as the CNS interaction only proceeds through a neutral current, it is equally sensitive to all active neutrino flavors. \\
\indent In Fig.~\ref{fig:eventrate} (left panel), we present all the neutrino fluxes that will induce relevant backgrounds to dark matter detection searches. The different neutrino sources considered in this study are the sun, which generates high fluxes of low energy neutrinos following the pp-chain \cite{ppneutrino} and the possible CNO cycle \cite{CNOneutrino1,CNOneutrino2}, diffuse supernovae (DSNB) \cite{DSNBneutrino} and the interaction of cosmic rays with the atmosphere \cite{atmosneutrino} which induces low fluxes of high energy neutrinos. As a summary of the neutrino sources used in the following, we present in Table~\ref{tab:neutrino} the different properties of the relevant neutrino families such as: the maximal neutrino energy, the maximum recoil energy for a Ge target nucleus and the overall flux normalization and uncertainty. In order to most directly compare to the analysis of Ref.~\cite{Billard:2013qya}, we use the standard solar model BS05(OP) and the predictions on the atmospheric  and the DSNB neutrino fluxes from \cite{atmosneutrino} and \cite{DSNBneutrino} respectively.\\

\begin{figure*}
\begin{center}
\includegraphics[width=0.95\columnwidth]{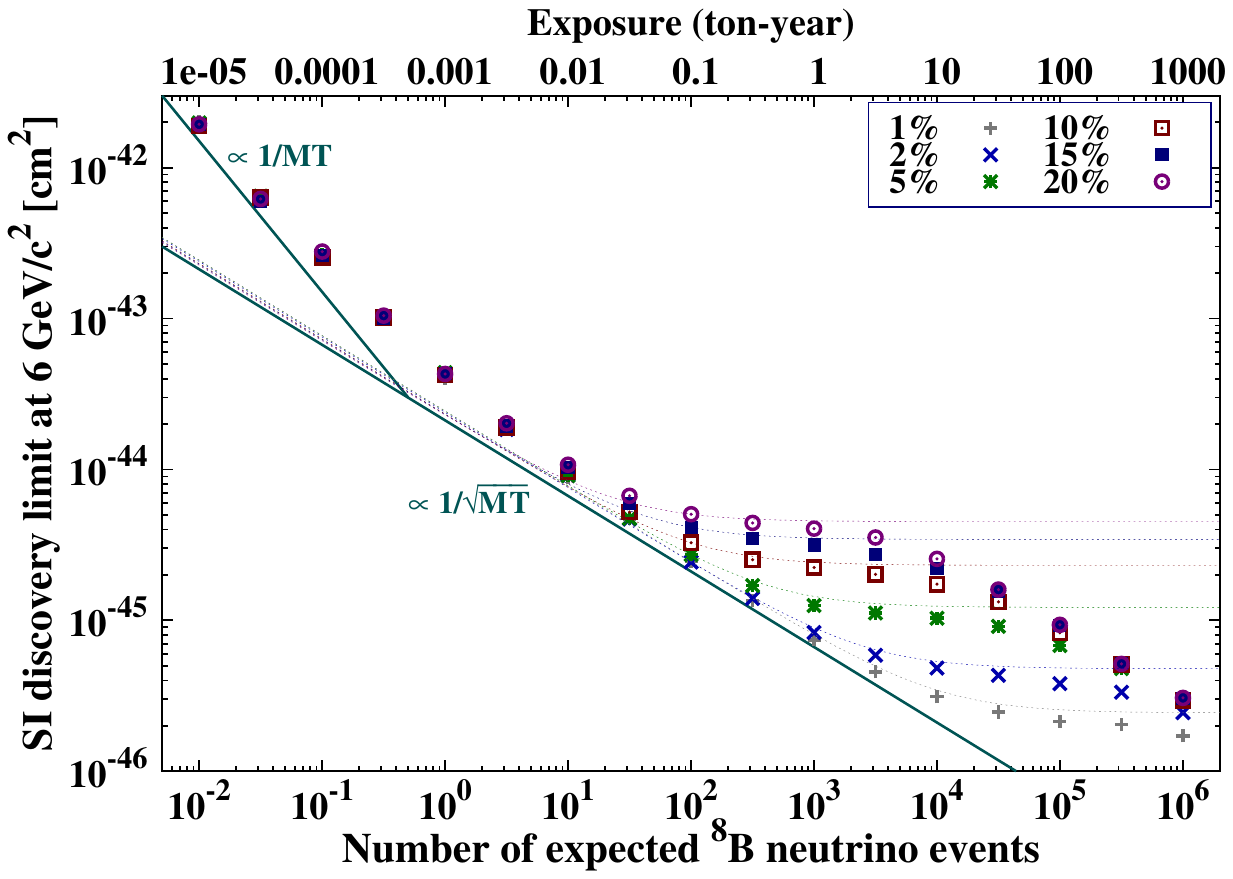}
\hspace{0.6cm}
\includegraphics[width=0.93\columnwidth]{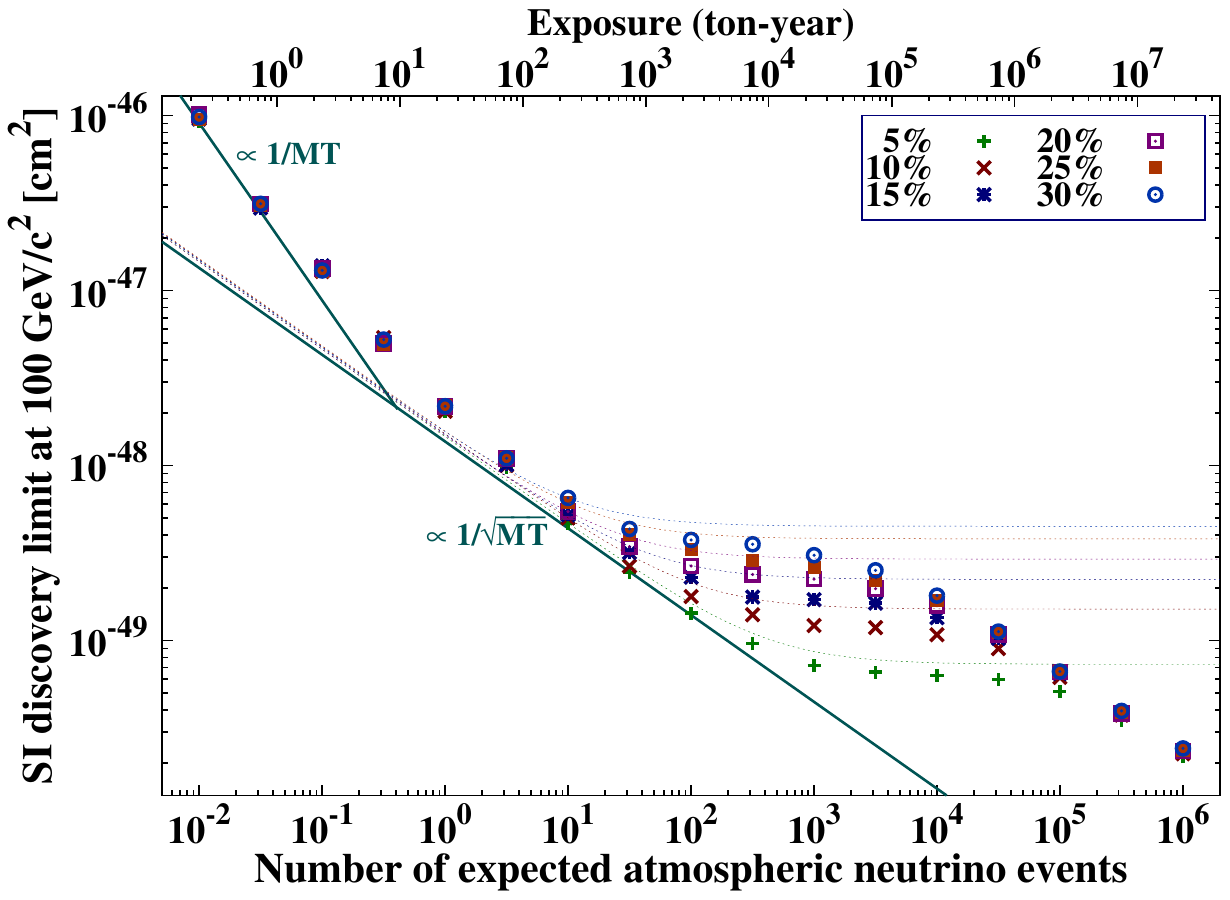}
\end{center}
\caption{Evolution of the discovery limit for a SI interaction as a function of the exposure for idealized Xe experiments with perfect efficiency and a 3~eV (4~keV) threshold. The discovery limit is shown for a $6~\rm{GeV/c^2}$ ($100~\rm{GeV/c^2}$) WIMP mass for different values of the systematic uncertainty on the ${}^{8}\rm{B}$ (atmospheric) flux in the left (right) panel. The second and third regions (background subtraction and saturation regime) are well described by equation \ref{eq:evol_disco} as shown by the dashed lines corresponding to the different systematic uncertainties.}
\label{fig:evolution_sys}
\end{figure*}

The different neutrino event rates are shown in Fig.~\ref{fig:eventrate} (right panel) for a Ge target. We can first notice that the highest event rates are due to the solar neutrinos and correspond to recoil energies below 6~keV. Indeed, the ${}^{8}\rm{B}$ and $hep$ neutrinos dominate the total neutrino event rate for recoil energies between 0.1 and 8 keV and  above these energies, the dominant component is the atmospheric neutrinos. Also shown, as a black solid line, is the event rate from a $6~\rm{GeV/c^2}$ WIMP with a SI cross section on the nucleon of $4.4\times 10^{-45}~\rm{cm^2}$. We can already notice that for this particular set of parameters $(m_{\chi},\sigma^{SI})$, the WIMP event rate is very similar to the one induced by the ${}^{8}\rm{B}$ neutrinos. As discussed in the next section, this similarity will lead to a strongly reduced discrimination power between the WIMP and the neutrino hypotheses and therefore dramatically affect the discovery potential of upcoming direct detection experiments.\\

Note that in this study we do not consider neutrino-electron scattering, even though it is predicted to provide a substantial signal in future dark matter detectors. Our primary motivation for this is because the neutrino-electron spectrum is flat and is therefore fairly easy to distinguish from a WIMP signal. Furthermore, in the following we will mainly focus on the low WIMP mass region (below $20~\rm{GeV/c^2}$) where the CNS background largely dominates over the neutrino-electron induced one. Moreover, most experiments are able to distinguish between electron and nuclear recoils down to 10$^{-3}$-10$^{-5}$, making the neutrino-electron scattering a negligible component.

\subsection{Discovery limit computation}
Following Ref.~\cite{Billard:2013qya}, we utilize a profile likelihood ratio test statistic in order to derive discovery limits of upcoming direct detection experiments in the context of the coherent neutrino scattering background. A discovery limit fixes a WIMP-nucleon cross section such that if the true WIMP-nucleon cross section is higher than this value then the considered experiment has a 90\% probability to detect a WIMP with at least a $3\sigma$ confidence level \cite{Discoverylim}. A binned likelihood function has been used in order to compute discovery limits for very high exposures:
\begin{align}
\mathscr{L}(\sigma_{\chi -n},\phi_{\nu}) &= \prod_{h=1}^{N_{exp}} \left[\prod_{i=1}^{N_{bin}}P\left(N^{h,i} \bigg{\vert} \mu_{\chi}^{h,i} + \sum_{j=1}^{N_{\nu}}\mu_{\nu}^{h,i,j}\right)\right] \nonumber \\& \times \prod_{j=1}^{N_{\nu}} \mathscr{L}_{\nu}^j(\phi_{\nu}^j) 
\label{likeli}
\end{align}
where $\phi_{\nu}^j$ are the different neutrino fluxes, $P$ is the Poisson probability function, $N_{exp}$ is the number of independent experiments, $N_{bin}$ is the considered number of bins, $N^{h,i}$ is the number of events in the $i$-th bin of $h$-th experiment and $N_{\nu}$ is the number of considered neutrino families. The values of $\mu_{\chi}^{h,i}$ and $\mu_{\nu}^{h,i,j}$ correspond respectively to the expected number of events from WIMPs and neutrinos of the family $j$ for the experiment $h$. They are computed by integrating the considered event rates over the recoil energy range of the $i$-th bin. Finally, $\mathscr{L}_{\nu}^j(\phi_{\nu}^j)$ are the individual likelihood functions related to the flux normalization of each neutrino component. They are parametrized as gaussian distributions with a standard deviation corresponding to the  uncertainty on the considered neutrino flux. \\

The profile likelihood ratio test statistic allows one to quantify the gap between a background only hypothesis ($H_ 0$) and an alternative hypothesis ($H_1$) which includes both background and signal \cite{profileratio}. It is defined as:
\begin{equation}
q_0 = \left\{ \begin{array}{l}
-2{\rm ln}\left[\frac{\mathscr{L}(\sigma_{\chi -n} = 0,\hat{\hat{\vec{\phi_{\nu}}}})}{\mathscr{L}(\hat{\sigma}_{\chi -n},\hat{\vec{\phi_{\nu}}})}\right]~~\hat{\sigma}_{\chi - n} > 0\\
~~~~~~~~~~~~~~~0~~~~~~~~~~~~~\hat{\sigma}_{\chi - n} < 0
\end{array}\right.
\end{equation}
\noindent where we used the $\hat{\hat{\vec{\phi_{\nu}}}}$ notation to show that this parameter varies in order to maximize the conditional likelihood function when $\sigma_{\chi -n}$ is fixed to zero. Following Wilk's theorem, the probability distribution function of $q_0$ asymptotically follows a half $\chi^2$ distribution with one degree of freedom. This has been checked by computing the histogram of the $q_0$ values under the $H_0$ hypothesis for 1,000 Monte Carlo pseudo-experiments. Therefore the significance of this test statistic is simply given by $Z = \sqrt{q_0^{obs}}$ in units of sigmas.\\

\begin{figure}[t]
\begin{center}
\includegraphics[width=0.93\columnwidth]{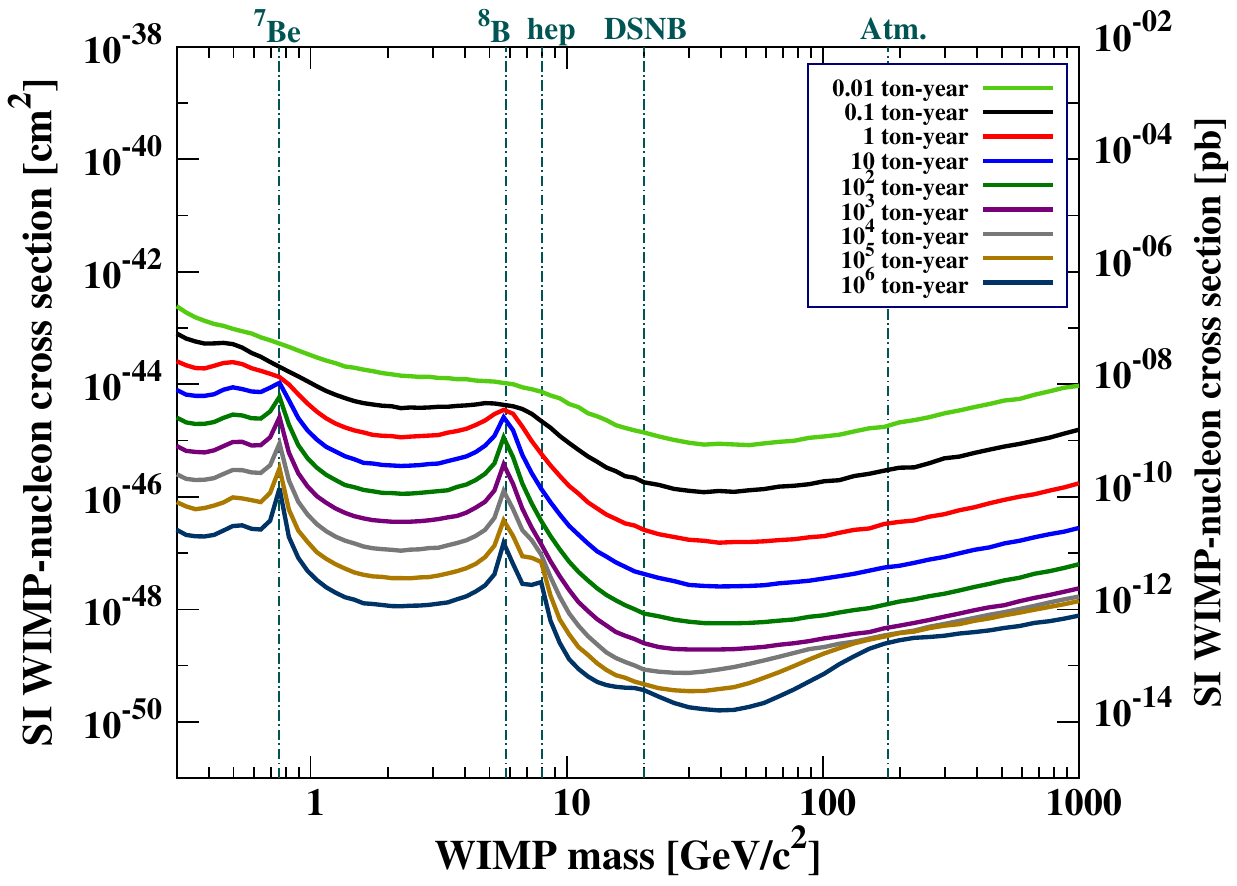}
\end{center}
\caption{Evolution of the discovery limit as a function of exposure on the WIMP mass vs SI cross section plane. These limits are computed for an idealized Xe-based experiment with no other backgrounds, 100\% efficiency, and an energy threshold of 3 eV to fully map the low WIMP mass discovery limit. Features appearing on the discovery limits with increasing exposures are due to the different components of the total neutrino background, see Table~\ref{tab:neutrino}.}
\label{fig:evolution_mass}
\end{figure}

Figure~\ref{fig:evolution_sys} presents the evolution of the discovery limit for a WIMP mass of 6 GeV/c$^2$ (left panel) and 100 GeV/c$^2$ (right panel) as a function of exposure which is given in number of expected neutrinos and ton-year on the bottom and top x-axes. Those calculations were done considering different values of the systematic uncertainties on the relevant neutrino background. From looking at Fig.~\ref{fig:evolution_sys}, one can understand the impact of the neutrino background on the discovery potential which is worth describing in a couple of main points:
\begin{itemize}
\item At the lowest exposures, where the neutrino background is negligible, and if no other backgrounds are present, the discovery potential evolves as $\sim 1/{MT}$ where $MT$ refers to the exposure (mass$\times$time) of the considered experiment.
\item As soon as the experiment starts to become sensitive to the neutrino background, the discovery potential evolves as $\sim 1/\sqrt{MT}$ as we are in a Poisson background subtraction regime.
\item When the exposure gets even bigger and the WIMP and neutrino signals are very similar, such as for a WIMP of 6 GeV/c$^2$ for a Xe target, the discovery potential starts to flatten out due to the systematic uncertainties. Indeed, in the extreme case where there is no discrimination power between the WIMP signal and the neutrino background, one would expect the discovery potential to evolve as \cite{Billard:2013qya}:
\begin{equation}
\sigma_{\rm{disco}} \propto \sqrt{\frac{1+\xi^2N_{\nu}}{N_{\nu}}}
\label{eq:evol_disco}
\end{equation}
where $\xi$ is the relative uncertainty on the relevant neutrino background. One can then see that the level and exposure at which the discovery potential flattens out is directly related to the level of this systematic uncertainty, as illustrated by Eq.~\ref{eq:evol_disco} and shown in Fig.~\ref{fig:evolution_sys}. This clearly highlights the need for reducing systematic uncertainties on the neutrino fluxes. Note that this saturation of the discovery potential can span about 2 orders of magnitude in exposure and therefore clearly represents a challenge to the development of future direct detection experiments.
\item Once enough neutrino events have been accumulated (between a few thousand and a million, depending on the systematic uncertainty), one can get back to a standard Poisson background subtraction regime and therefore overcome the previously described saturation regime. This is due to the small differences in the tails of the neutrino- and WIMP-induced spectra which lead to additional discrimination power, e.g., Fig~\ref{fig:eventrate} (right panel) around $6.5~\rm{keV}$ for the 6 GeV/c$^2$ case. However, these small differences in the induced spectra only become relevant at very high exposures (especially at the high WIMP mass region) which are well beyond what is envisioned for the next generation of experiments. 
\end{itemize}

In Figure~\ref{fig:evolution_mass} we illustrate the evolution of the discovery limit, in the light of the neutrino background, as a function of the WIMP mass. We use an idealized Xe-based experiment with a recoil energy threshold of 3~eV and perfect efficiency to map out the low and high WIMP mass range. This figure clearly shows that there are some particular WIMP mass ranges for which we expect the neutrino background to dramatically affect the discovery potential, {\it i.e.} going through a saturation regime. A few examples are for masses around $m_\chi$ = 0.8 GeV/c$^2$, $m_\chi$ = 6 GeV/c$^2$, $m_\chi$ = 8 GeV/c$^2$ and above $m_\chi$ = 100 GeV/c$^2$ where the WIMP signal is well mimicked by the $^7$Be, $^8$B, $hep$ and the atmospheric neutrinos respectively. For WIMP masses with strong differences between the WIMP and the neutrino recoil spectra, the discovery potential evolves close to $1/\sqrt{MT}$ as one can see from Fig.~\ref{fig:evolution_mass} for WIMP masses between 1 GeV/c$^2$ and 4 GeV/c$^2$ for example.

\begin{table*}
\begin{ruledtabular}
\begin{tabular}{|cccccc|}
Target & Sample Experiment & $\mathbf{E_{\rm th}^{\rm{low}}}$ \bf{(eV)} & $\mathbf{E_{\rm th}^{\rm{high}}}$ \bf{(keV)}  & \bf{Exposure}$\mathbf{^{\rm{low}}}$ (ton-yr) & \bf{Exposure}$\mathbf{^{\rm{high}}}$ ($\times10^3$ ton-yr)\\
\hline
Xe & LZ/XENON1T & 3 & 4 & 0.19 & 9.3\\
Ge & SuperCDMS/CoGeNT & 5.3 & 7.9 & 0.38 & 15.6\\
Si &SuperCDMS/DAMIC &  14 & 20 & 1.26 & 73.1\\
Ar & DEAP/DarkSide & 9.6 & 14.4 & 0.72 & 32.5\\
CaWO$_4$ & CRESST & 25 & 35 & 1.48 & 24.4\\
C$_{3}$F$_{8}$ & PICO & 33 & 47.7 & 2.02 & 25.1\\
CF$_{4}$ & MIMAC/DMTPC & 33 & 47.7 & 2.39 & 22.9\\
CF$_3$I & PICO/COUPP & 33 & 47.7 & 2.42 & 23.8\\
\end{tabular}
\caption{Corresponding energy thresholds and exposures for each target material used to derive the discovery limit over the low and the high WIMP mass regions. The thresholds are defined such that for the low WIMP mass region one we expect no $pp$ neutrino events in the data and that for the high WIMP mass region one we expect no ${}^{8}\rm{B}$ neutrinos events in the data. The exposures are set such that the low WIMP mass regime detects 200 events from ${}^{8}\rm{B}$ (about 1660 neutrino events total including pep, hep and ${}^{7}\rm{Be}$), and the high WIMP mass regime detects 400 neutrino events.\label{tab:Exposures}}
\end{ruledtabular} 
\end{table*}

Interestingly, from Figs.~\ref{fig:evolution_sys} and \ref{fig:evolution_mass} we can clearly see that the neutrino background will impact the low WIMP mass (below 10 GeV/c$^2$) and the high WIMP mass (above 10 GeV/c$^2$) regions at very different exposures due to the vastly different rates of neutrino backgrounds seen in Fig~\ref{fig:eventrate}. Indeed, the discovery limit evolution with exposure transitions from 1/MT to 1/$\sqrt{MT}$ as soon as the expected neutrino background nears 1 event, which corresponds to exposures of a few kg-years for very-low-threshold experiments in the low-mass region and a few tens of ton-years for the high-mass region for idealized perfect-efficiency experiments (actual exposure values depend on the target, threshold, and efficiency of a given experiment). The saturation regime exists for exposures between 100 kg-year and 10 ton-year and between 10$^3$ ton-year and 10$^6$ ton-year for the low and high WIMP mass regions respectively. Therefore, one can conclude that the next generation of low-threshold experiments focusing on the low WIMP mass region could reach well into the $1/\sqrt{MT}$ regime and even begin to see the effects of saturation, while at high WIMP masses the next generation of experiments will at most begin to see the 1/$\sqrt{MT}$ effects from the neutrino background.

\subsection{Discovery limits for different targets}

From Fig.~\ref{fig:evolution_sys} and \ref{fig:evolution_mass}, we have seen that the neutrino background can lead to a saturation regime of the discovery potential over certain mass ranges that can span over 2 orders of magnitude in exposure. However, with enough exposure or lower systematics, one can always improve the discovery limit. In the following, we will define an arbitrary neutrino induced discovery limit within the saturation regime, derived from a set of two different exposures (and energy thresholds) for the low and high WIMP mass regimes. The thresholds are defined such that for the low mass region we expect no $pp$ neutrino events in the data and that for the high mass region we expect no ${}^{8}\rm{B}$ neutrinos events in the data. The exposures are set such that the low WIMP mass regime detects 200 events from ${}^{8}\rm{B}$ (about 1660 neutrino events total including pep, hep and ${}^{7}\rm{Be}$), and the high WIMP mass regime detects 400 neutrino events. The corresponding exposures and thresholds for each target discussed in the following are summarized in Table~\ref{tab:Exposures}. One can then see, using figure \ref{fig:evolution_sys}, that the computed limits are always within the saturation regime of the discovery potential for both the low and the high WIMP mass regions. As previously discussed, this is where adding a new observable such as annual modulation~\cite{modulation} and/or directionality~\cite{directionality_Ahlen,directionality_Grothaus}, or combining data from several experiments, as suggested in Sec.~\ref{sec:beyond}, can lead to the most substantial improvement in the discovery potential.

\begin{figure*}
\begin{center}
\includegraphics[width=0.95\columnwidth]{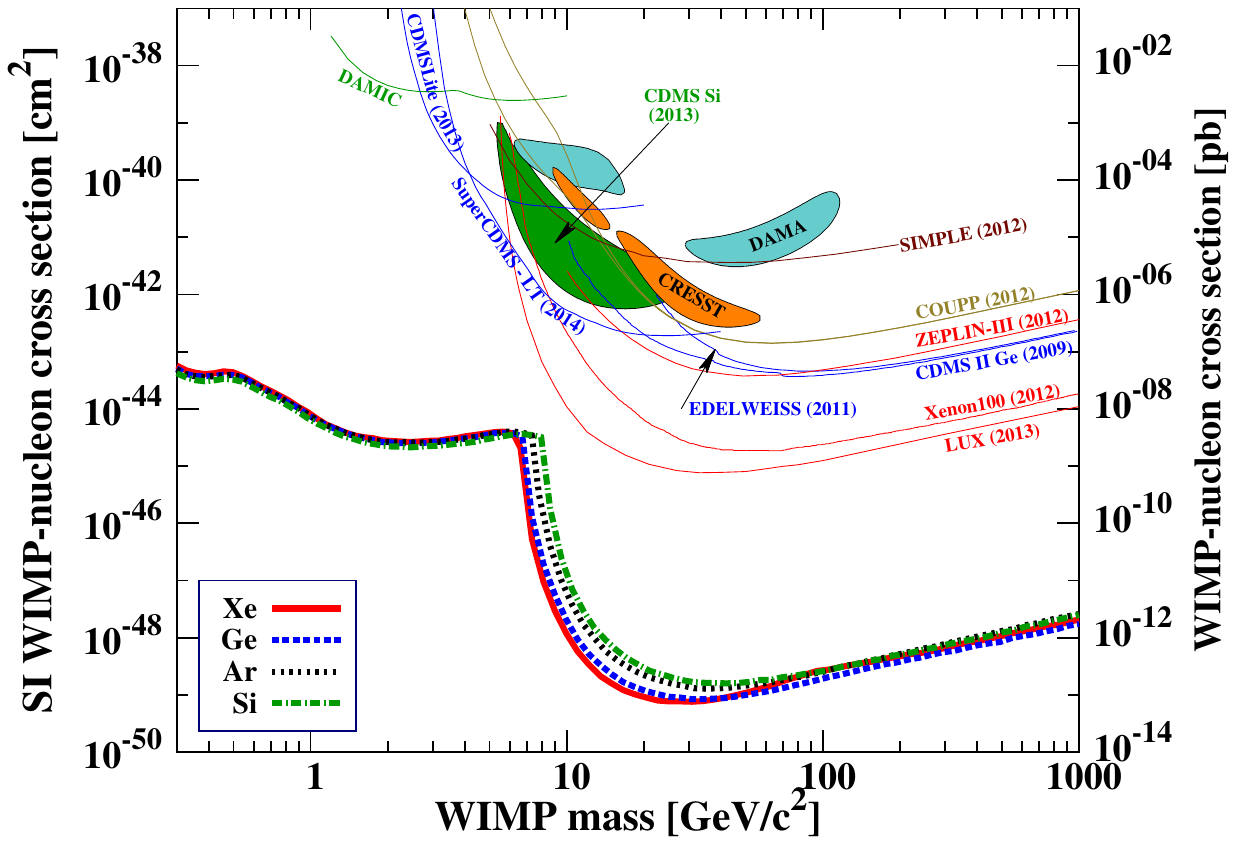}
\hspace{0.5cm}
\includegraphics[width=0.95\columnwidth]{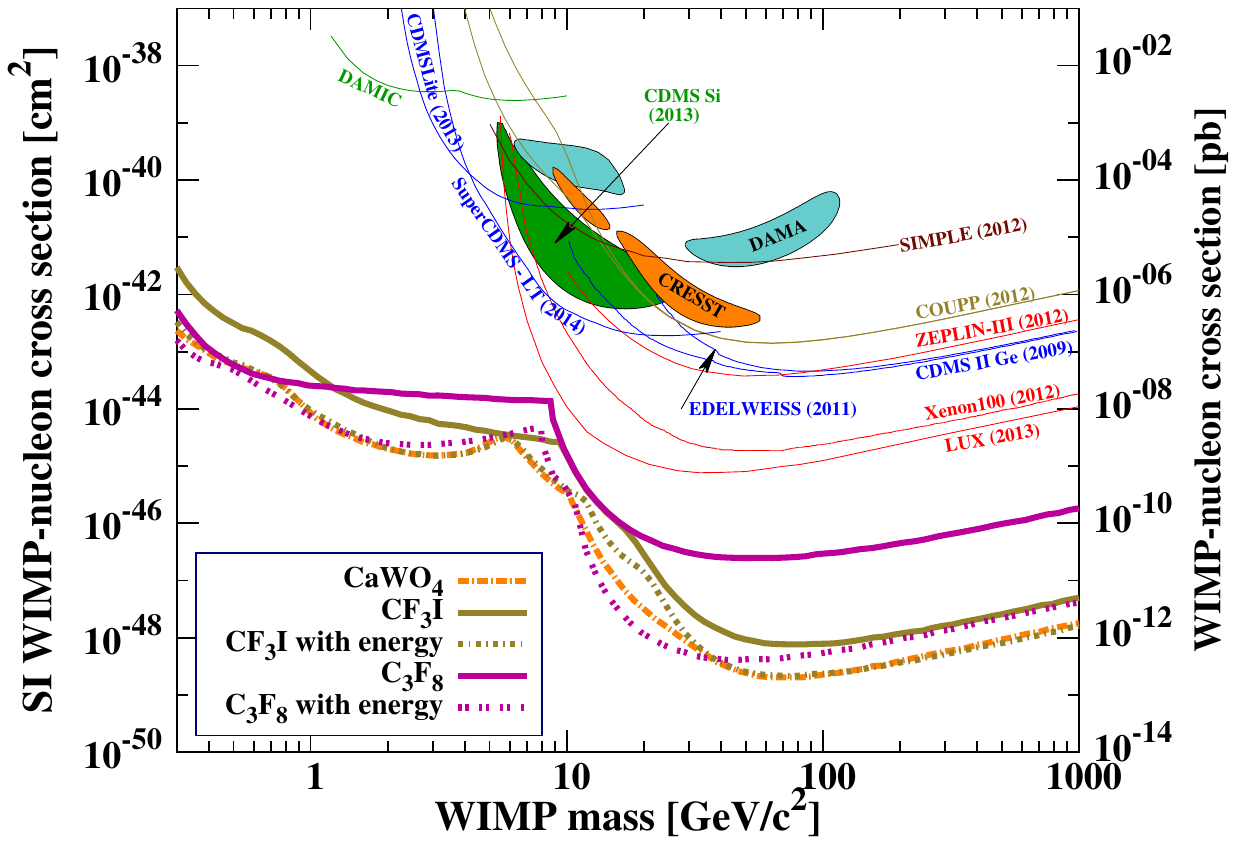}
\end{center}
\caption{Discovery limits for single (left) and multi target (right) based experiments considering the SI interaction. Also shown are the current sensitivity limits of several experiments: DAMIC \cite{DAMIC}, SIMPLE \cite{SIMPLE}, SuperCDMS-LT \cite{SuperCDMSLT}, COUPP \cite{COUPP}, ZEPLIN-III \cite{ZEPLIN3}, EDELWEISS \cite{EDELWEISS}, CDMS II Ge \cite{CDMS2}, CDMSLite \cite{CDMSLite}, XENON100 \cite{XENON100}, LUX \cite{LZ}, CDMS-II Si \cite{CDMSSi}, DAMA/LIBRA \cite{DAMA} and CRESST \cite{CRESST}.}
\label{fig:discovery_limit_SI}
\end{figure*}

\begin{figure*}
\begin{center}
\includegraphics[width=0.95\columnwidth]{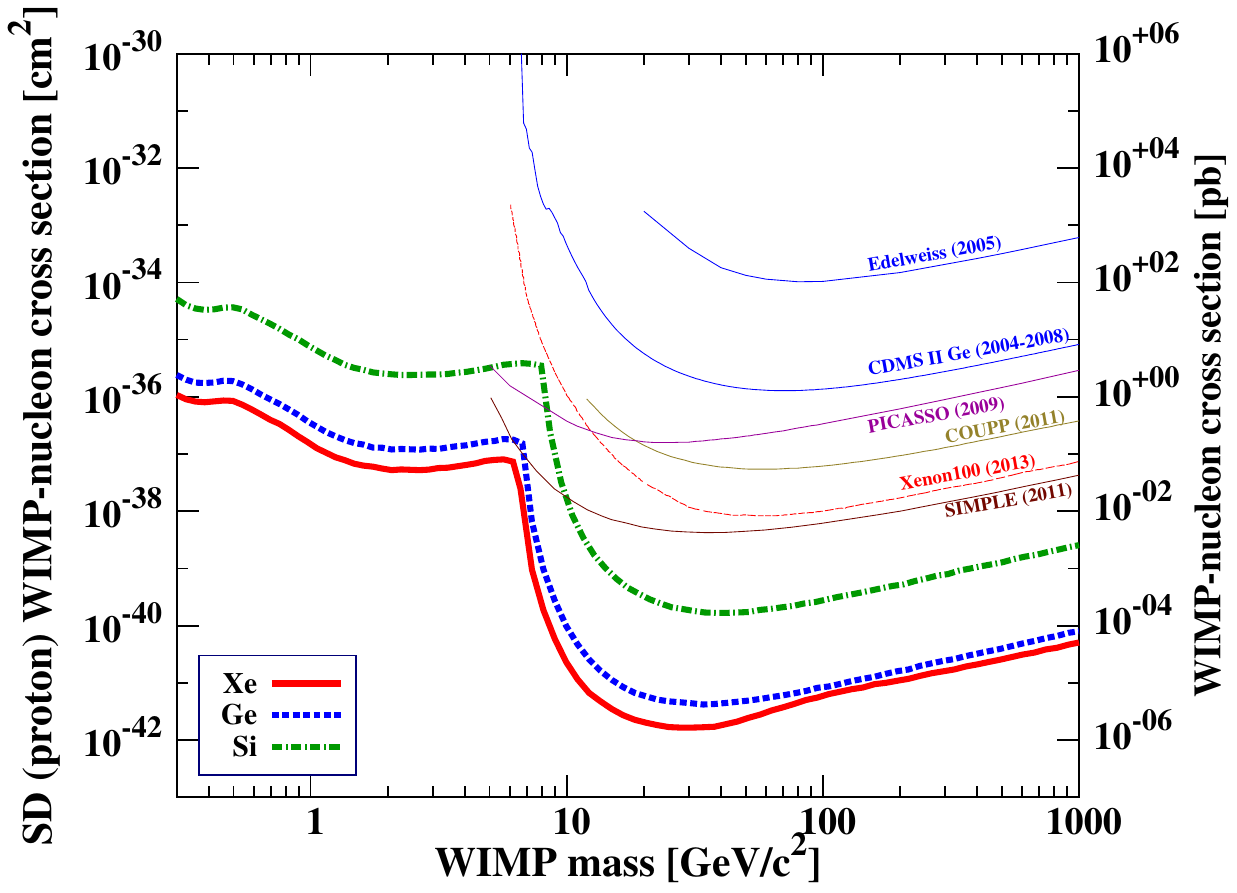}
\hspace{0.5cm}
\includegraphics[width=0.95\columnwidth]{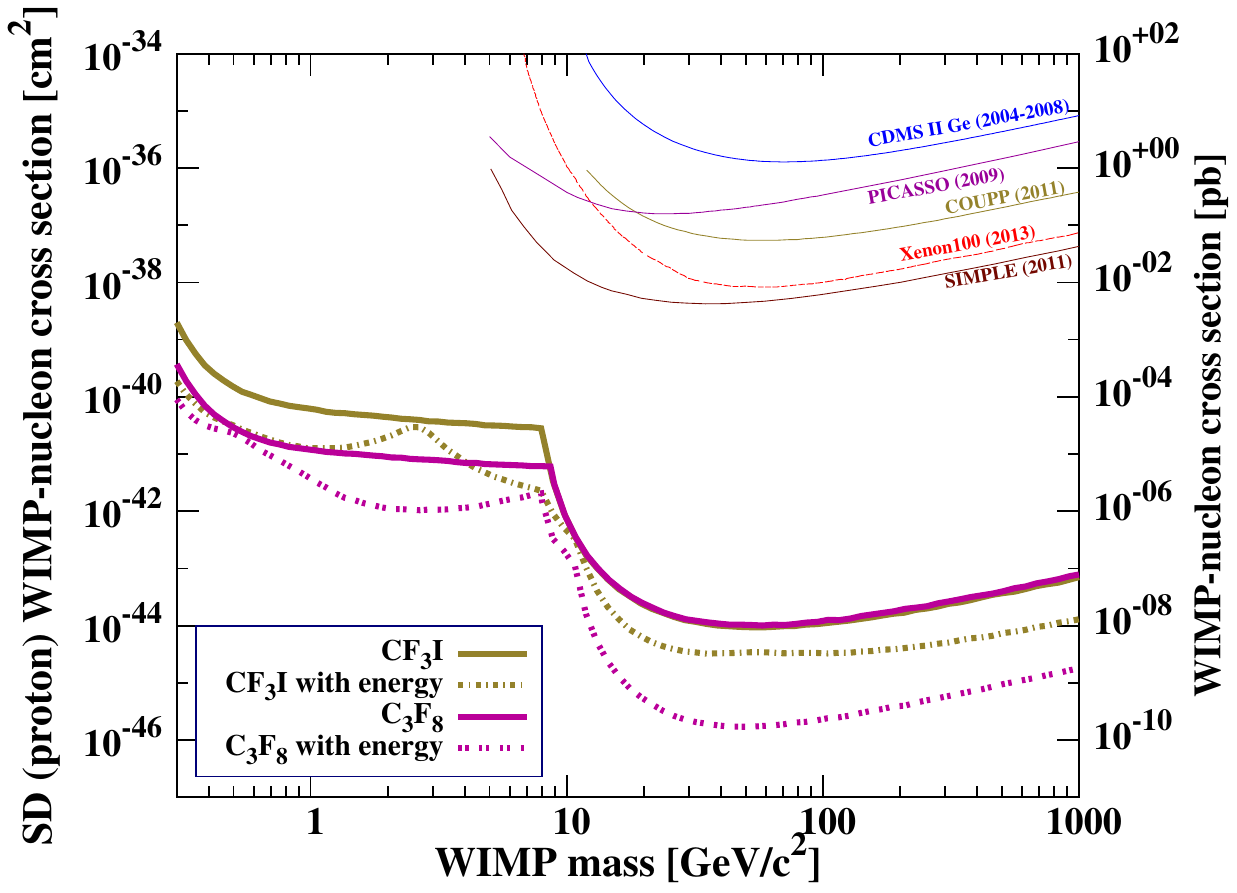}
\end{center}
\caption{Same as figure~\ref{fig:discovery_limit_SI} but considering the SD interaction on the proton.}
\label{fig:discovery_limit_SDp}
\end{figure*}

\begin{figure*}
\begin{center}
\includegraphics[width=0.95\columnwidth]{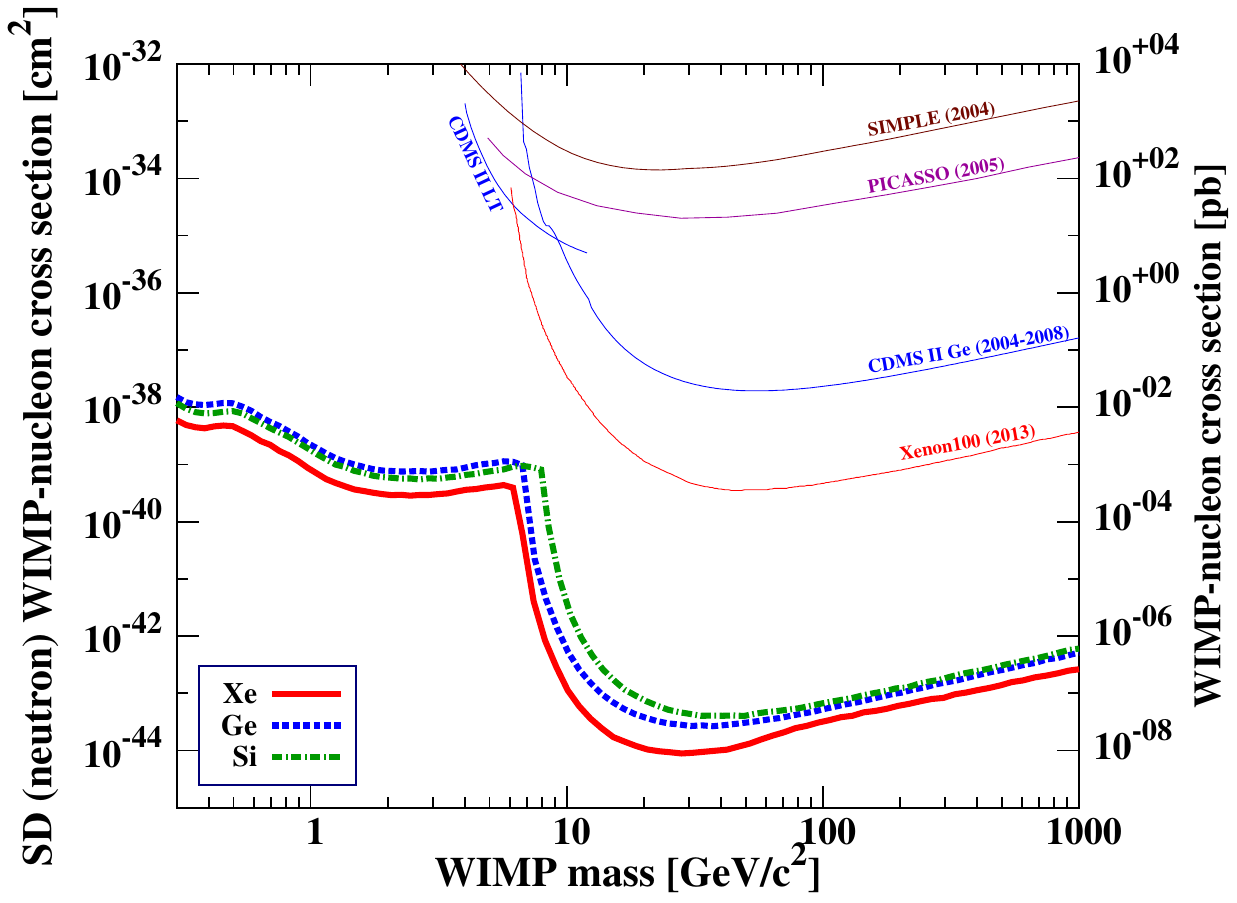}
\hspace{0.5cm}
\includegraphics[width=0.95\columnwidth]{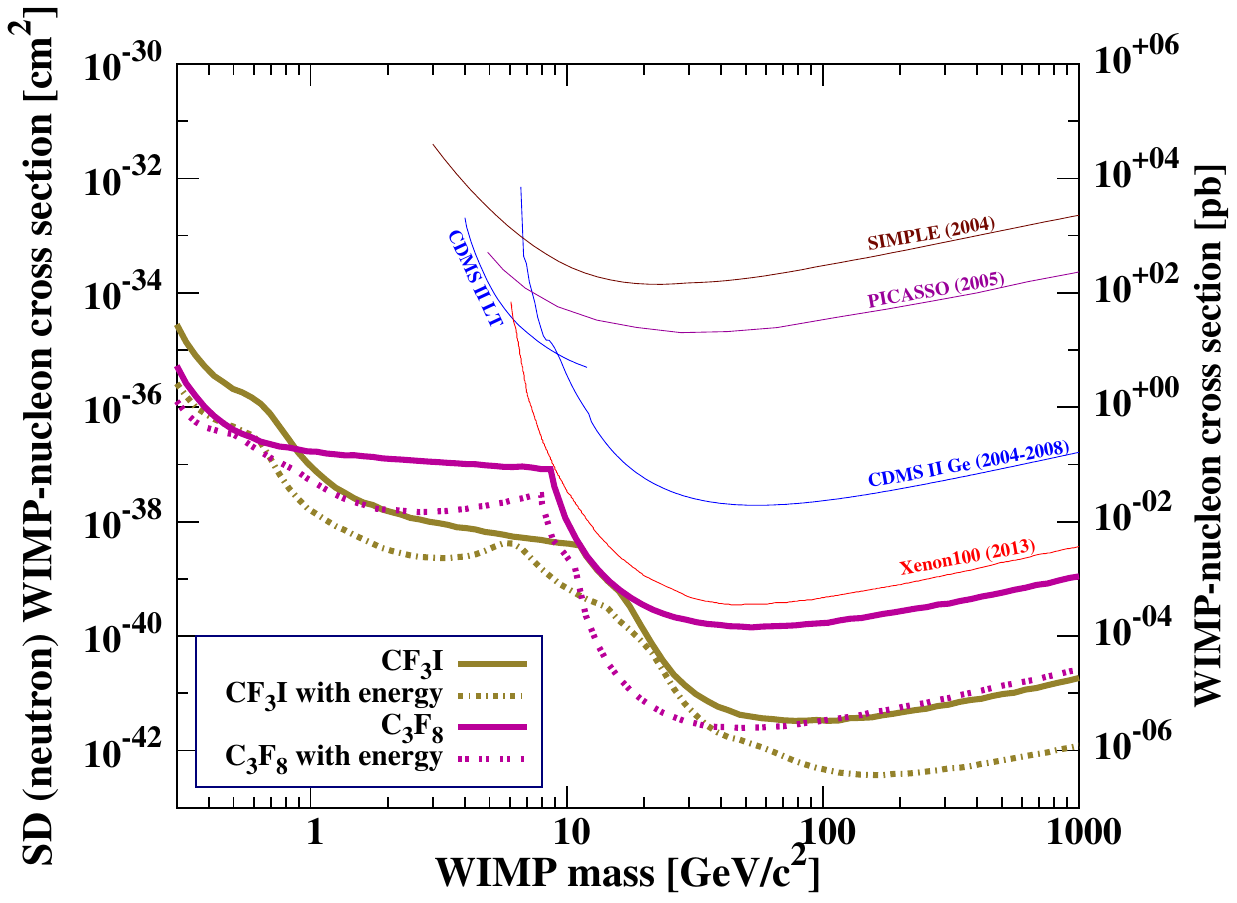}
\end{center}
\caption{Same as figure~\ref{fig:discovery_limit_SI} but considering the SD interaction on the neutron.}
\label{fig:discovery_limit_SDn}
\end{figure*}

Figure \ref{fig:discovery_limit_SI}, \ref{fig:discovery_limit_SDp} and \ref{fig:discovery_limit_SDn} show the computed limits for single-target and multi-target-based experiments for both SI and SD interactions using the target and neutrino properties shown in tables \ref{tab:nuclei} and \ref{tab:neutrino}, and the energy thresholds and exposures from table~\ref{tab:Exposures}. It should be noted that these are limits calculated for idealized experiments with perfect efficiency, and no background except for neutrinos, and thus represent the best attainable discovery limit for the listed exposures.

In figure \ref{fig:discovery_limit_SI} (left panel) we show the discovery limits of elemental targets considering only the SI interaction. One can first notice that, using a binned likelihood function, we have been able to reproduce the results from \cite{Billard:2013qya}. In the SI case the equivalent WIMP models corresponding to a given neutrino type are only weakly dependent on the considered target (see Sec.~\ref{sec:beyond} for more details), thus all single-nucleus based targets share a very similar discovery limit.

On figure \ref{fig:discovery_limit_SI} (right panel) we show the discovery limits for the SI interaction considering compound targets such as $\rm{CaWO_4}$ (used in CRESST), $\rm{CF_3I}$ (used in COUPP), and $\rm{C_3F_8}$ (used in PICO and PICASSO). Since  experiments that use $\rm{C_3F_8}$ and $\rm{CF_3I}$ don't currently measure the recoil energy of the event, we computed limits with and without energy sensitivity, shown as dashed and solid lines respectively. In the case of compound targets, we chose the thresholds by considering the lightest nucleus of the mixture as it has its neutrino spectra the most shifted to the high recoil energies. For example, considering $\rm{CaWO_4}$, in order to compute the low threshold part of the discovery limit, we chose a threshold such that there is no pp neutrino events in the data due to the O target, \emph{i.e} 25~eV. By choosing this threshold we do not consider the same event rate proportions for the three different targets. This explains the three successive exponential-like falls from a WIMP mass of $6~\rm{GeV/c^2}$ to about $100~\rm{GeV/c^2}$ for this target. We can apply the same reasoning for the $\rm{CF_3I}$ target where we can see two different exponential-like falls due to I and F targets. For $\rm{C_3F_8}$ this effect is not visible as F and C have very similar atomic masses. Note that the discovery limit for CF$_4$ (used in experiments such as MIMAC~\cite{Santos:2013hpa} and DMTPC~\cite{Lopez:2013ah}) is almost identical to the one from a C$_3$F$_8$ based experiment with energy sensitivity. Therefore, all the following results using $\rm{C_3F_8}$ are also relevant to CF$_4$.

\begin{figure*}
\begin{center}
\includegraphics[width=0.95\columnwidth]{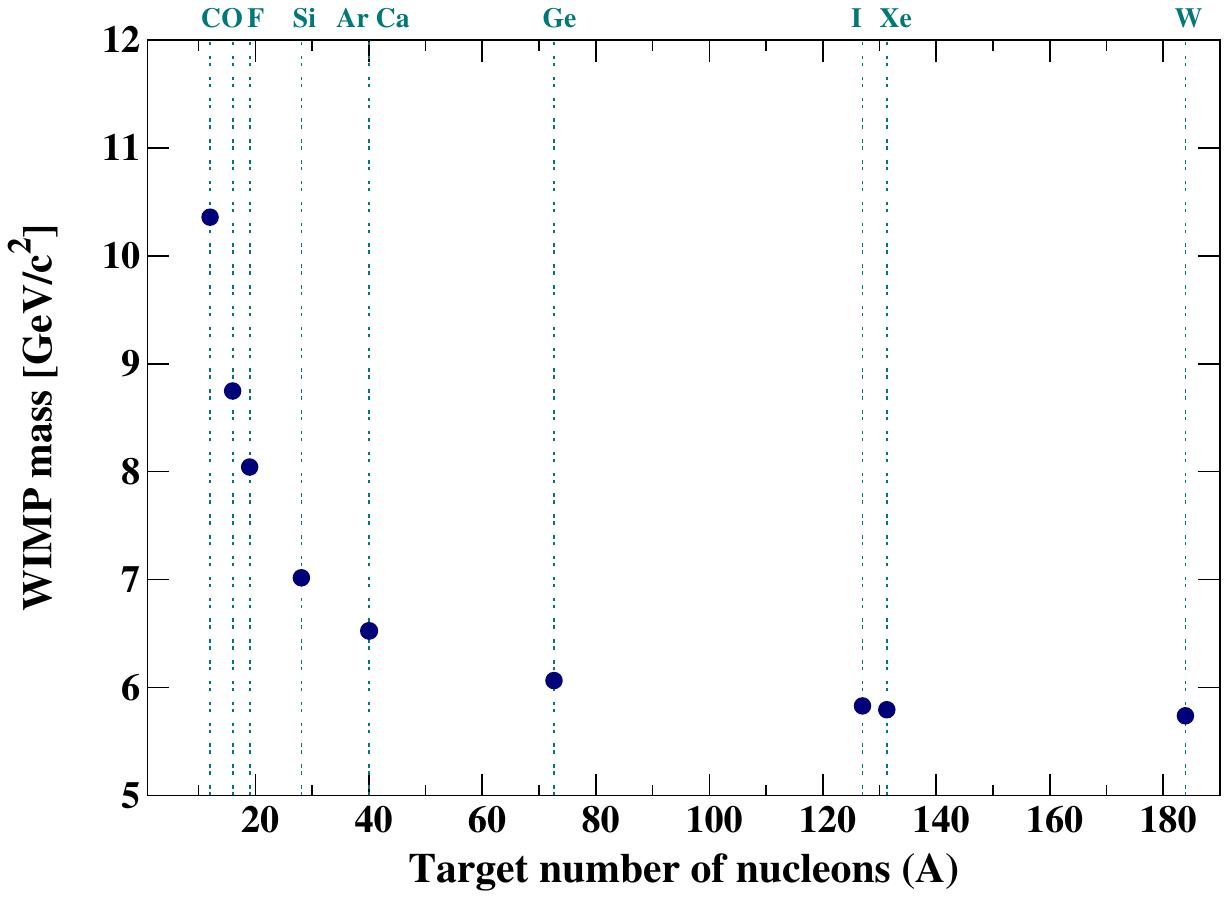}
\hspace{0.5cm}
\includegraphics[width=0.95\columnwidth]{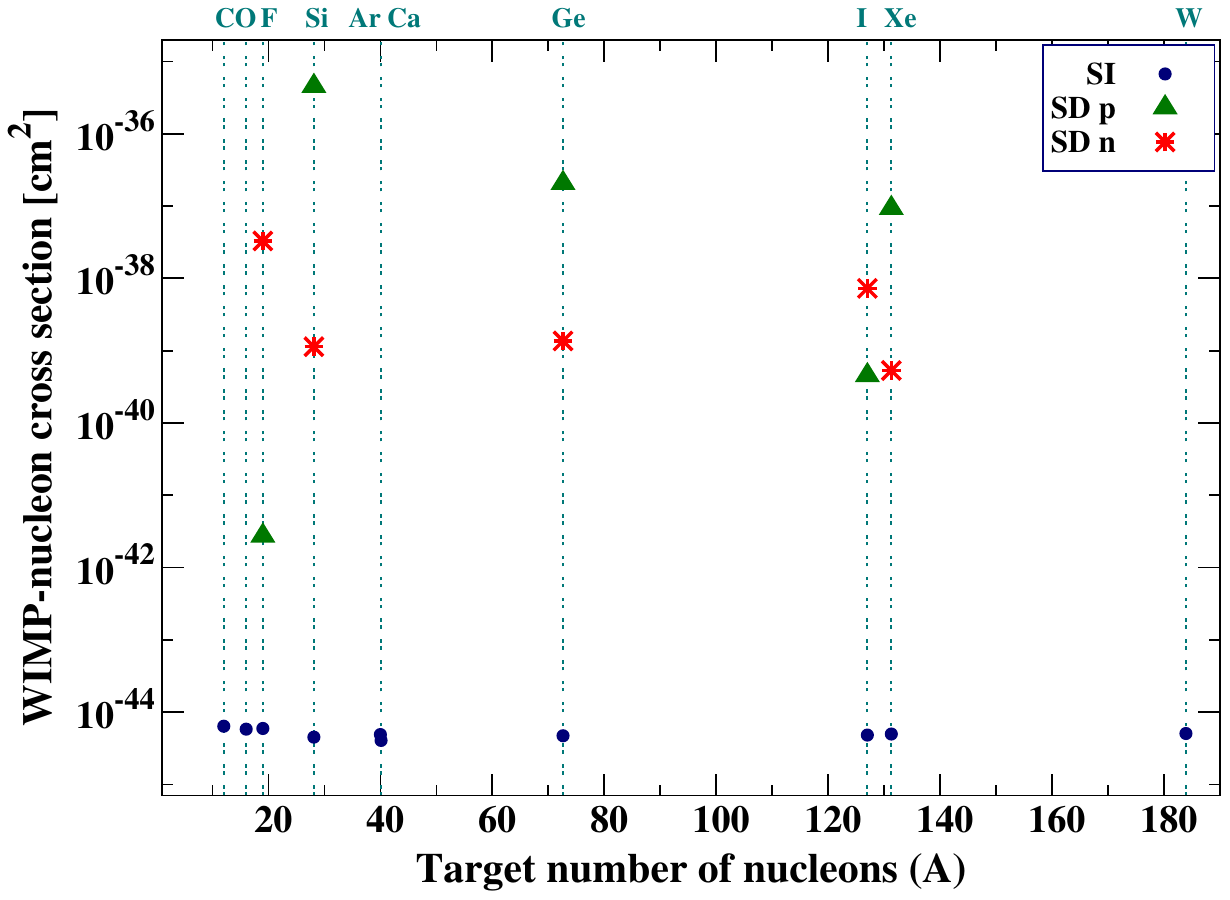}
\end{center}
\caption{Left: reconstructed WIMP-masses when considering the ${}^{8}\rm{B}$ neutrino background under the WIMP only hypothesis. Right: figure of merit of target complementarity showing the mean values of the reconstructed WIMP-nucleon cross sections computed by considering the distributions of the maximum likelihood of the ${}^{8}\rm{B}$ neutrino under the WIMP only hypothesis for different target nuclei and 1,000 Monte Carlo pseudo-experiments.}
\label{fig:complementarity}
\end{figure*}

Figures~\ref{fig:discovery_limit_SDp} and \ref{fig:discovery_limit_SDn} show the discovery limits considering a SD interaction on the proton and on the neutron, respectively. These discovery limits have similar shapes to those derived for the SI interaction but their relative amplitudes can be very different. For example, considering a SD interaction on the proton, a Si-based experiment and an experiment using $\rm{C_3F_8}$ have their discovery limits around $10^{-35}~\rm{cm^2}$ and $10^{-41}~\rm{cm^2}$ in the low mass region, respectively. This difference is due to the nature of the SD interaction. Indeed, the neutrino-nucleus cross section is still described by a coherent effect (evolution in $\rm{A^2}$) while the SD WIMP-nucleus cross section depends on the total angular momentum and mean spin contents of the considered target (see eq. \ref{eq:sigSD} and table \ref{tab:nuclei}). Another interesting feature that one can notice from figures \ref{fig:discovery_limit_SDp} and \ref{fig:discovery_limit_SDn} concerns the multi-target based experiments. Indeed, for such experiments, some nuclei from the target material cannot lead to a SD interaction as they have $J = 0$.  This effect leads to different shapes in the discovery limits between the SI and SD cases, as one can see for $\rm{C_3F_8}$ where $J_{C} = 0$ and $J_{F} = 1/2$. One may also notice that $\rm{CaWO_4}$ does not appear on figures \ref{fig:discovery_limit_SDp} and \ref{fig:discovery_limit_SDn} as it has no SD sensitivity.

For these six figures, the discovery limit indicates the region of parameter space where the neutrino background has its highest impact on the reach of upcoming direct detection experiments. Indeed, with the considered energy thresholds and exposures, the discovery potential is in the saturation regime over most of the mass range, where a slight increase in sensitivity would be at the cost of a large increase in exposure. This clearly highlights the need for developing alternative strategies to be able to probe dark matter models lying underneath this neutrino-induced bound on the WIMP discovery reach.

\section{Combining data from different experiments}
\label{sec:beyond}

As shown in the previous section, it can be extremely difficult to claim a discovery of dark matter if the true WIMP model lies below the neutrino background. However, there are several possible ways to improve the discovery potential, even in the neutrino-induced saturation regime. The first one is to improve the theoretical estimates and experimental measurements of neutrino fluxes as it has been shown in figure \ref{fig:evolution_sys}. A second possibility is to add some new  observables that could help at disentangling between the WIMP and neutrino origin of the observed nuclear recoils. This could be done by searching for annual modulation~\cite{modulation} and/or by measuring the nuclear recoil direction~\cite{directionality_Grothaus}, as suggested by upcoming directional detection experiments \cite{directionality_Ahlen}. Indeed, since the main neutrino background has a Solar origin, the directional signature of such events is expected to be drastically different from the WIMP induced one \cite{directionality_Billard_2010,directionality_Billard_2011}.

\subsection{Target complementarity}

\begin{figure*}
\begin{center}
\includegraphics[width=0.95\columnwidth]{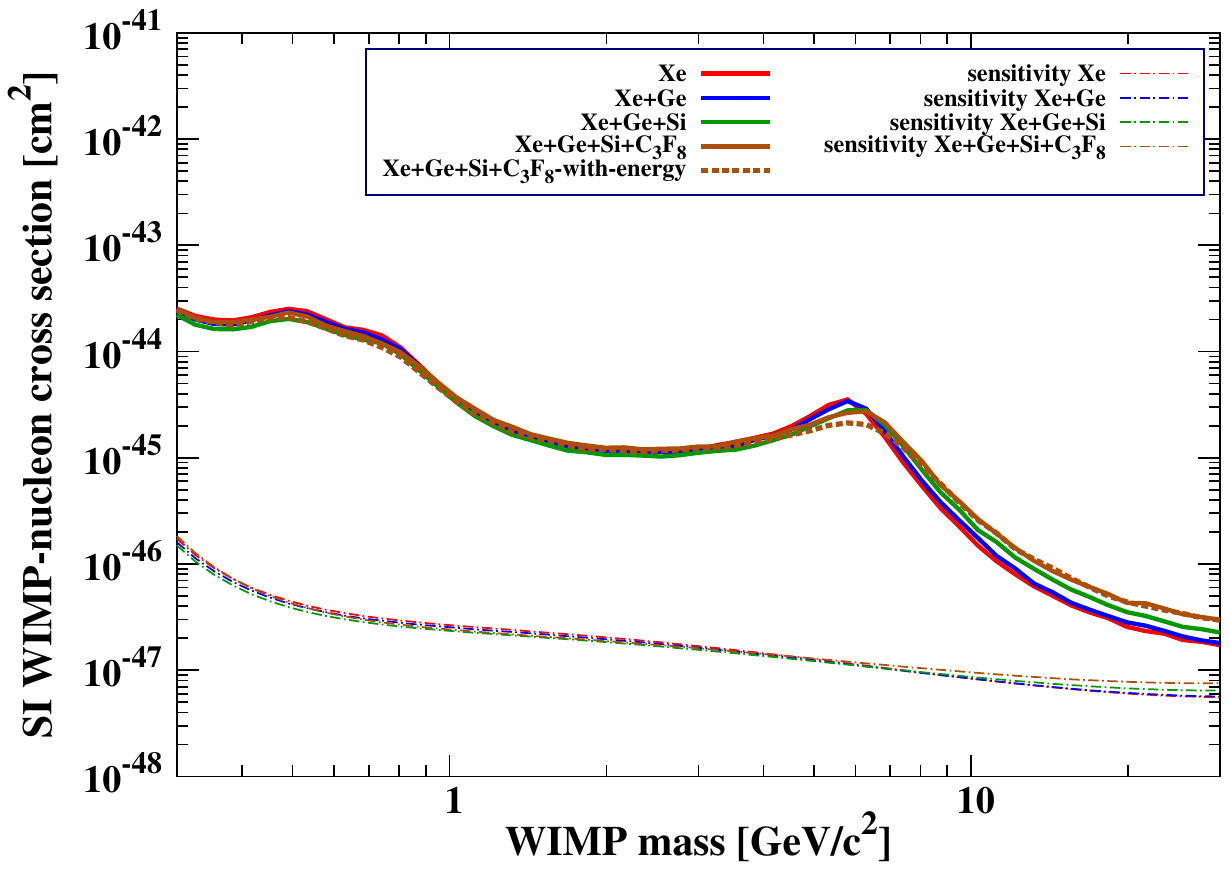}
\hspace{0.5cm}
\includegraphics[width=0.95\columnwidth]{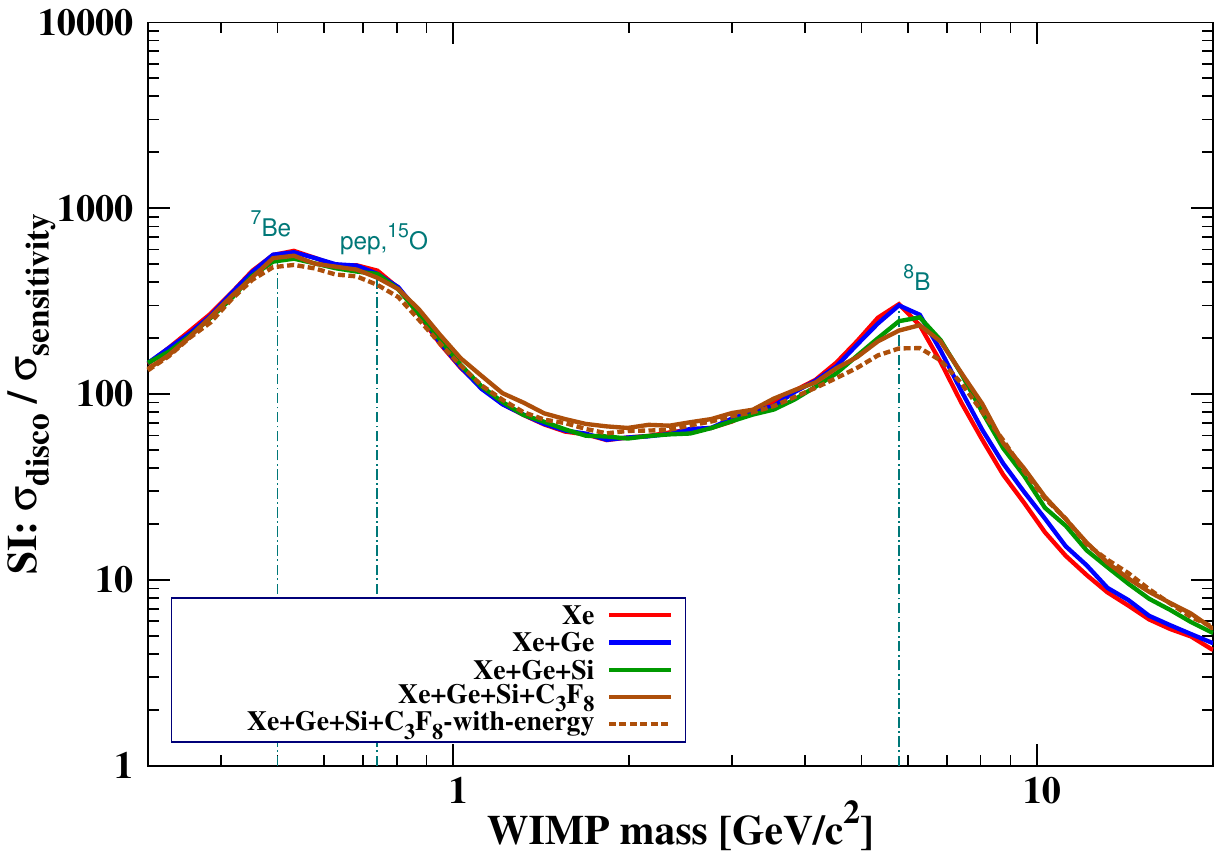}
\end{center}
\caption{Left: Discovery limits for a Xe only, and combinations of Xe/Ge, Xe/Ge/Si and Xe/Ge/Si/$\rm{C_3F_8}$ based experiments for the SI interaction with thresholds set such that we do not have pp neutrino events in the data: 3~eV, 5.3~eV, 14~eV and 33~eV for Xe, Ge, Si and C respectively. The exposure has been set such as an expected number of 1,000 ${}^{8}\rm{B}$ neutrino events equally distributed between the targets was expected in the four cases. For the $\rm{C_3F_8}$ based experiment we computed two limits with and without energy sensitivity. Also shown are the background-free limits obtained without considering the neutrino background (sensitivity limits). Right: Ratio between the discovery limits of a Xe based experiment, combinations of Xe/Ge, Xe/Ge/Si and Xe/Ge/Si/$\rm{C_3F_8}$ based experiments with their respective sensitivity limits. The gap between this ratio and $1$ gives the effect of the background on the discovery potential. Also shown as dashed vertical lines the WIMP masses that can be mimicked by different neutrino components.}
\label{fig:complementarity_SI}
\end{figure*}

\begin{figure*}
\begin{center}
\includegraphics[width=0.95\columnwidth]{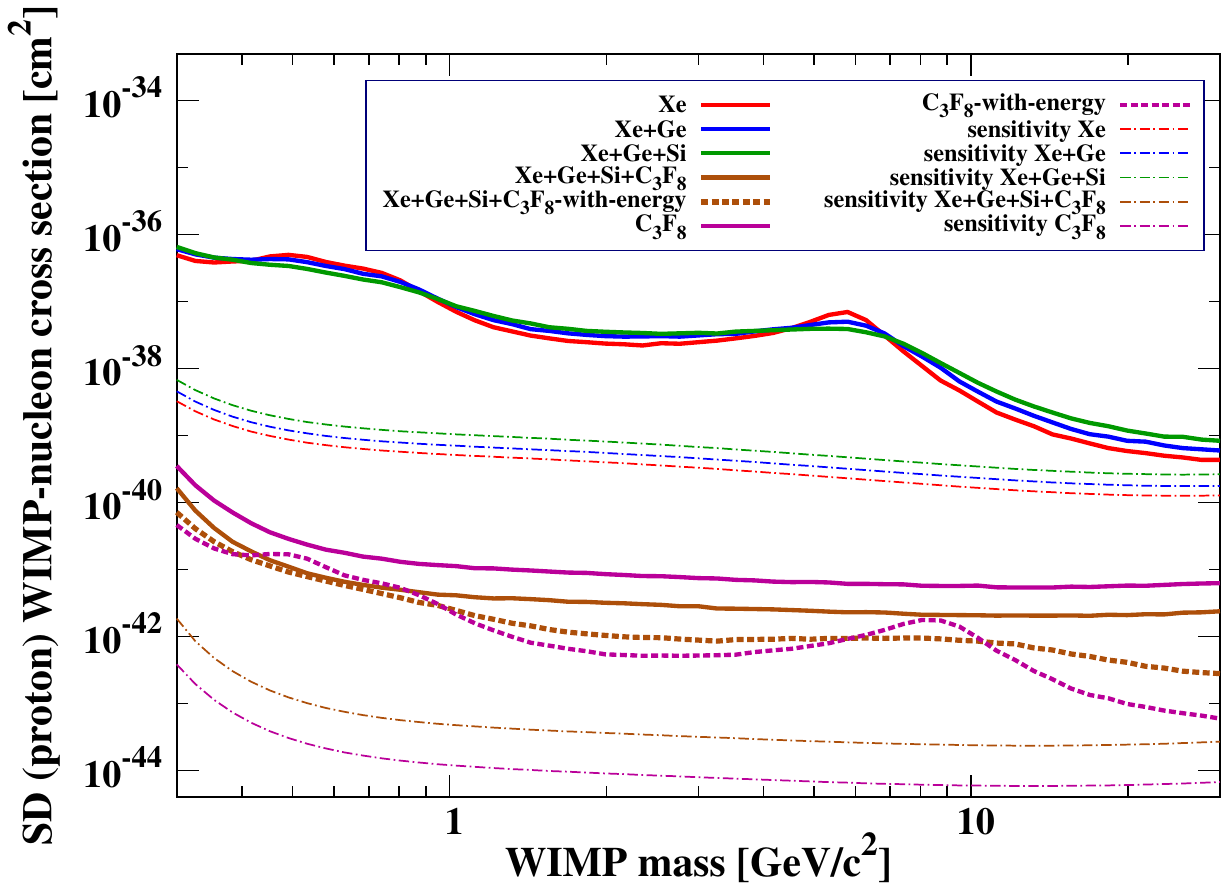}
\hspace{0.5cm}
\includegraphics[width=0.95\columnwidth]{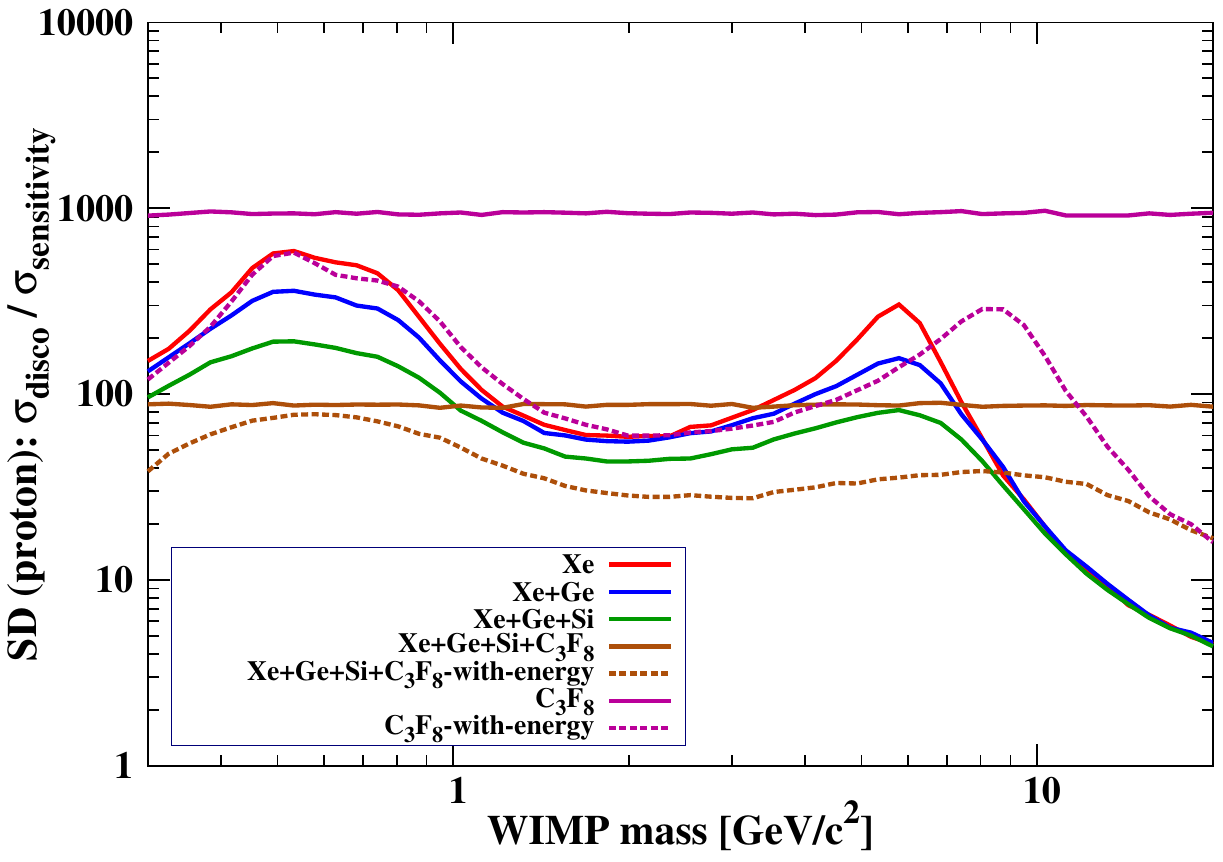}
\end{center}
\caption{Same as figure~\ref{fig:complementarity_SI} but considering the SD interaction on the proton. For the $\rm{C_3F_8}$ based experiment we computed two limits with and without energy sensitivity. The right panel shows clearly that combining data from $\rm{C_3F_8}$ with other targets results in a significant decrease of the impact of the neutrino background on the discovery potential.}
\label{fig:complementarity_SDp}
\end{figure*}

\begin{figure*}
\begin{center}
\includegraphics[width=0.95\columnwidth]{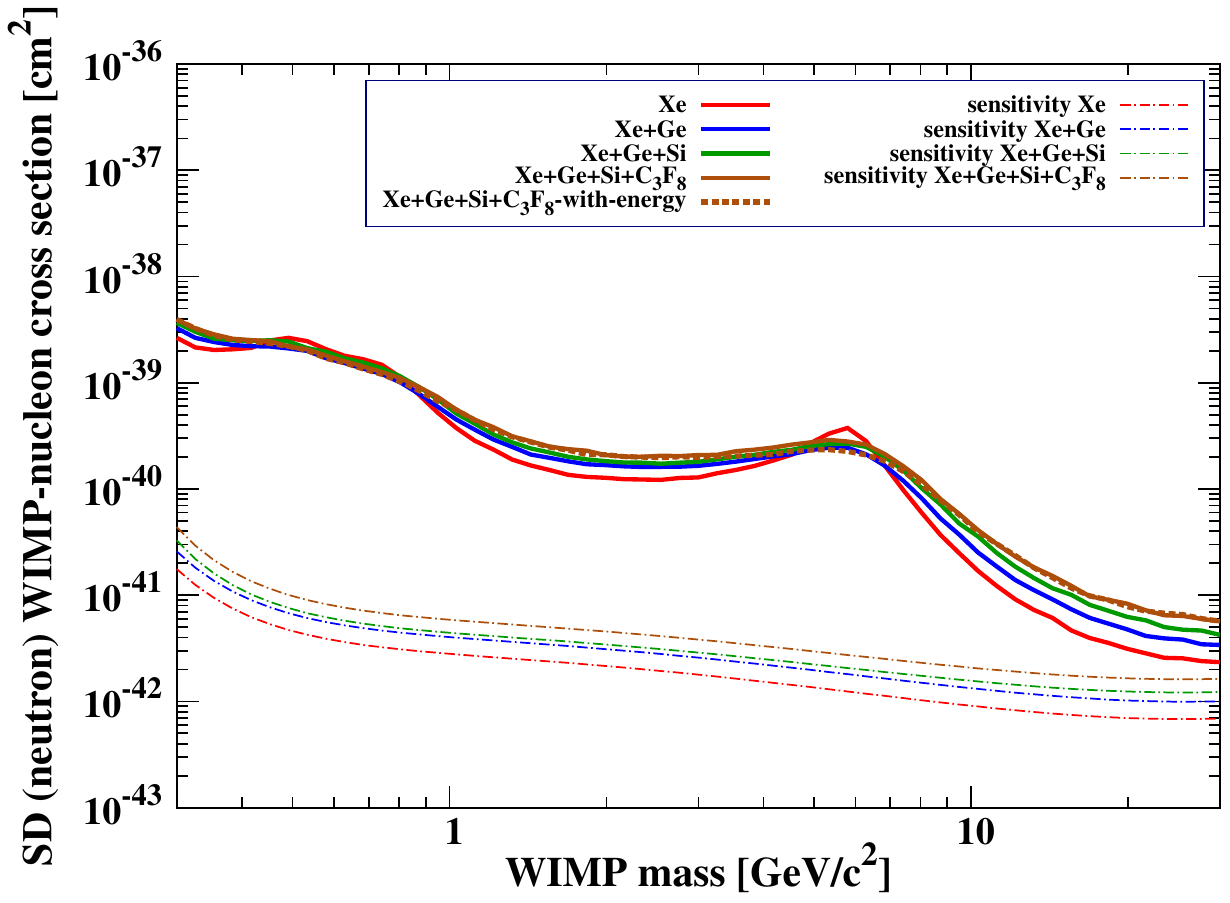}
\hspace{0.5cm}
\includegraphics[width=0.95\columnwidth]{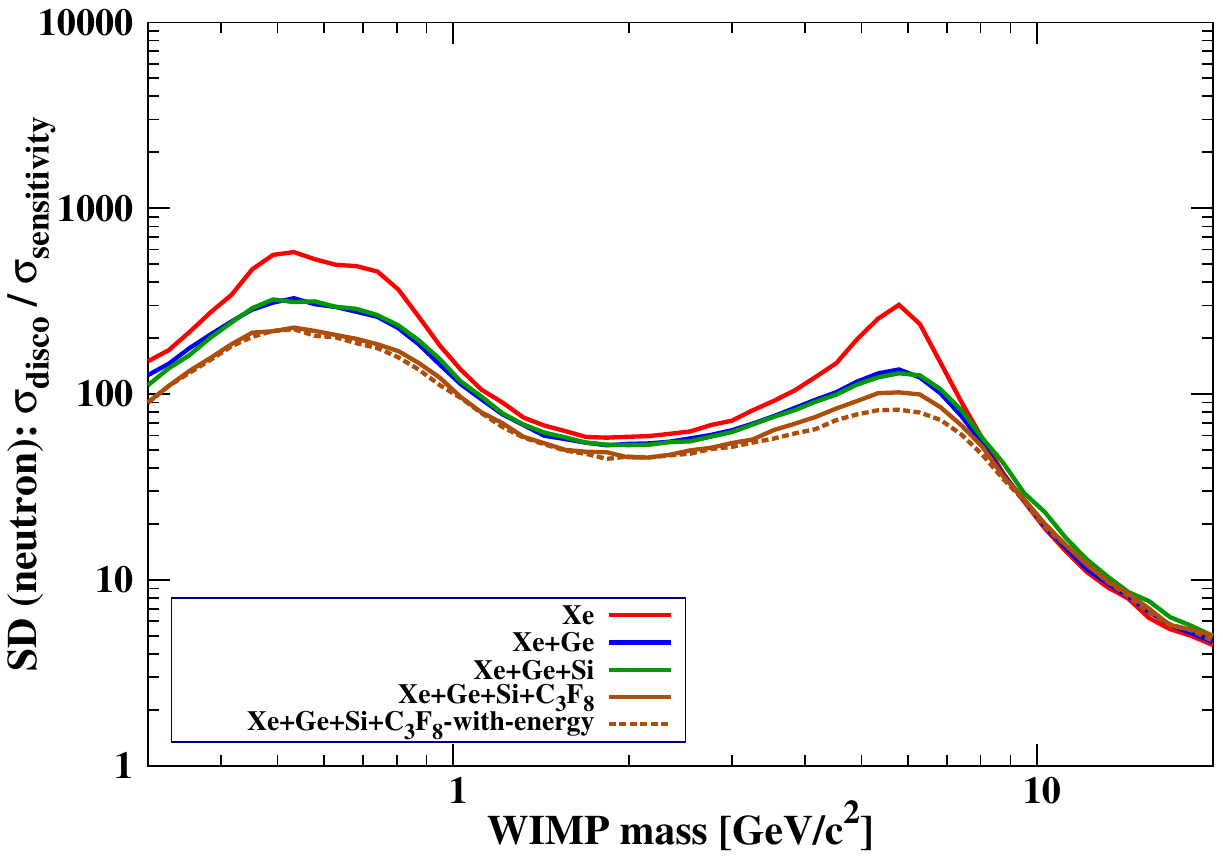}
\end{center}
\caption{Same as figure~\ref{fig:complementarity_SI} but considering the SD interaction on the neutron.}
\label{fig:complementarity_SDn}
\end{figure*}

We are now in position to study the gains in discovery potential that can be achieved by combining data from different target materials in light of the neutrino background. As discussed in the previous section, the effect of neutrino background is notably important at particular WIMP masses and cross sections where the WIMP and neutrino spectra are similar, both in shape and magnitude. The WIMP mass and cross section where this occurs is target-material dependent. One may then wonder whether combining different experiments could alleviate such degeneracies between the WIMP and the neutrino hypotheses.

In order to investigate this possibility, we determine what WIMP models can be mimicked by $^8$B neutrinos by computing the maximum likelihood distributions under the WIMP only hypothesis in the ($m_\chi$, $\sigma^{SI,SD}$) plane from fake data containing only nuclear recoils from $^8$B neutrinos. Because of the small differences between the WIMP and neutrino energy distributions, the reconstructed WIMP masses for each target are slightly different. This is result shown in figure \ref{fig:complementarity} (left panel). 
For a SI interaction, the reconstructed cross sections for different targets are all roughly the same as is shown on figure \ref{fig:complementarity} (right panel).
This is because both neutrino- and WIMP-nucleus cross sections evolve as $A^2$ (Eq. \ref{eq:sigSI}). Therefore, we expect the target complementarity in the SI case to be fairly weak. 

However, for a SD interaction the situation is different. As figure~\ref{fig:complementarity} (right panel) clearly indicates, the reconstructed cross sections for the different targets can be different by many orders of magnitude. This is due to the nature of the WIMP SD cross section which depends on the angular momentum ($J$), spin content ($\langle S_{p,n}\rangle$) and isotopic fractions ($f_A$) of the considered targets (Eq. \ref{eq:sigSD}). Thus, as the reconstructed WIMP models of a given neutrino for different target nuclei are strongly distinguishable, one could expect to gain in discrimination power between the WIMP and the neutrino hypothesis by combining data from several experiments.  As a matter of fact, figure~\ref{fig:complementarity} (right panel) can be seen as a way to quantify the target complementarity. Indeed, the larger the difference in reconstructed WIMP cross sections between two targets is, the greater the discrimination power and improvement in a combined discovery potential becomes.

\begin{figure*}
\begin{center}
\includegraphics[width=0.95\columnwidth]{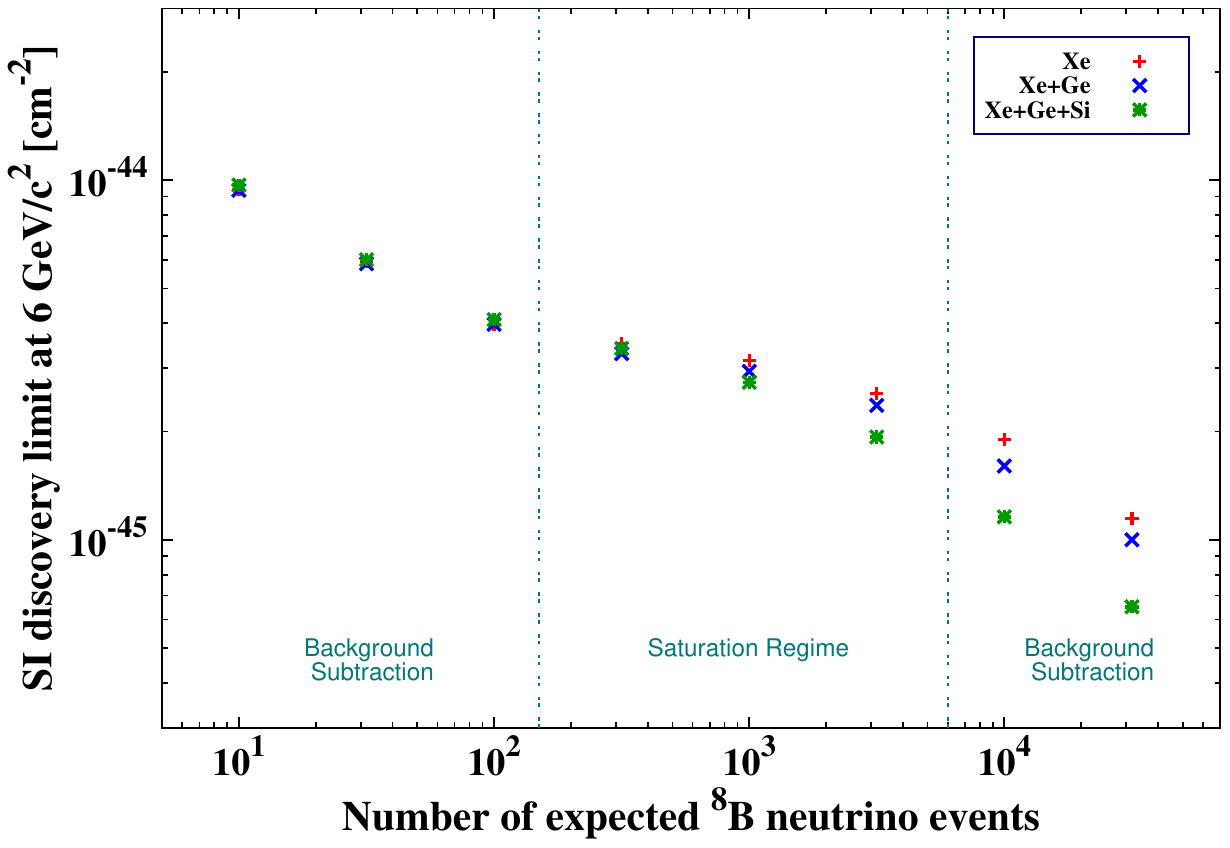}
\hspace{0.5cm}
\includegraphics[width=0.95\columnwidth]{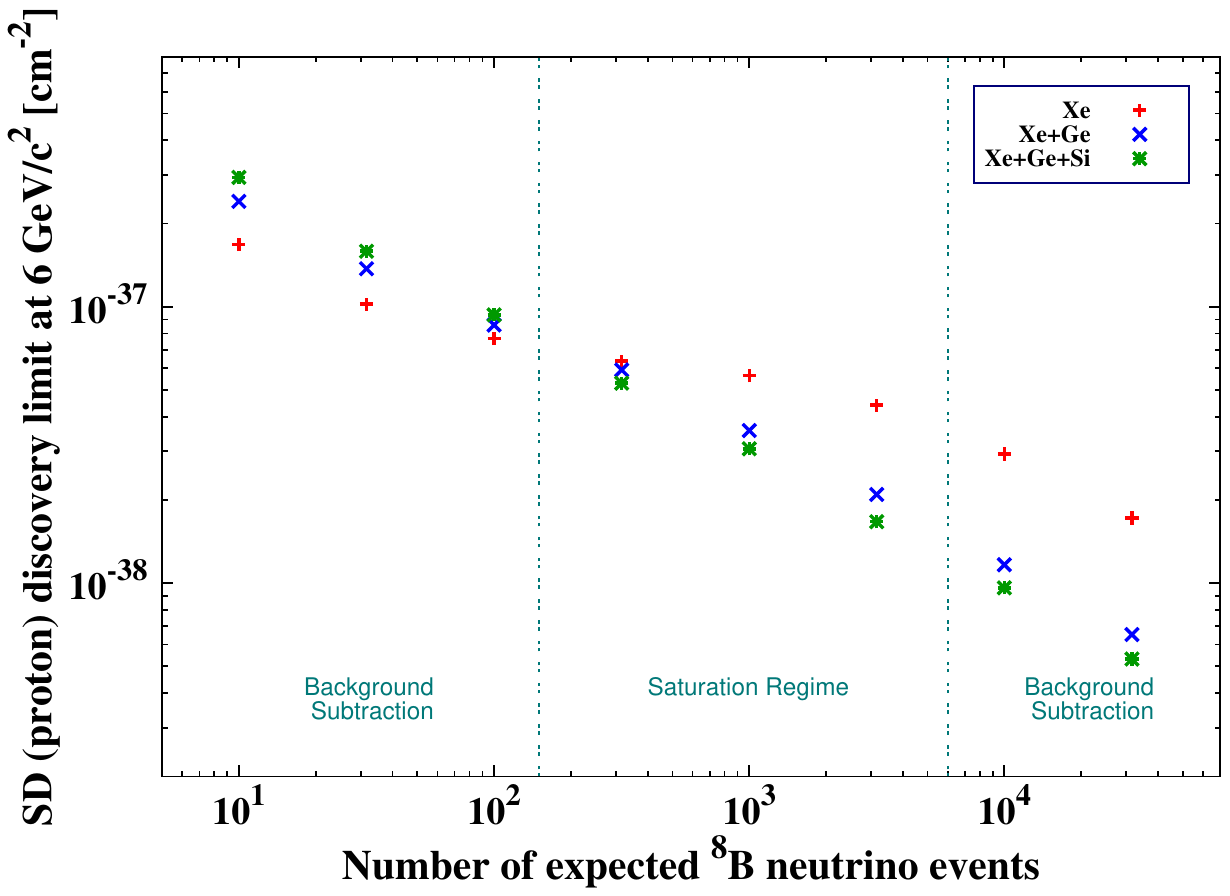}
\end{center}
\caption{Evolution of the discovery limit as a function of the exposure for a Xe based experiment and combinations of  Xe/Ge, Xe/Ge/Si based experiment for a $6~\rm{GeV/C^2}$ WIMP mass. The considered threshold for Xe, Ge, and Si are respectively 1.0~keV, 1.8~keV, and 4.7~keV. We considered both the SI interaction (left panel) and the SD interaction on the proton (right panel). For an expected number of 100 ${}^{8}\rm{B}$ events, the corresponding exposures for Xe, Xe/Ge, and Xe/Ge/Si are respectively 1.005 ton-year, 1.989 ton-year, and 3.154 ton-year.}
\label{fig:evolution}
\end{figure*}

As mentioned above, target complementarity is not very efficient for the SI interaction as one can see on figure~\ref{fig:complementarity_SI} (left) where we have computed the discovery limits (solid curves) for progressive combinations of Xe, Ge, Si, and $\rm{C_3F_8}$ targets using thresholds taken from Table~\ref{tab:Exposures} for the low WIMP mass region. The exposures for each individual target have been set such that we expect a total of 1,000 ${}^{8}\rm{B}$ neutrino events equally distributed amongst each experiment. For example, the Xe+Ge curve has exposures set such that 500 events are detected in Xe and 500 in Ge.
The slight shift to higher WIMP masses when adding Si to the Xe+Ge mixture comes from the differences in the reconstructed WIMP masses from a $^8$B signal shown in Fig.~\ref{fig:complementarity} (left panel).  $\rm{C_3F_8}$ is currently used in bubble chamber experiments such as COUPP which do not have sensitivity to the recoil energy since they detect events which deposit an energy density above the threshold required to generate bubble nucleation \cite{COUPP}. We have thus included curves with and without energy sensitivity for this target. For reference, the background-free sensitivity limits for the chosen exposures are also shown as dashed lines. The ratio between the sensitivity and the discovery limits, shown in Fig.~\ref{fig:complementarity_SI} (right panel), allows us to quantify the impact of the neutrino background on the WIMP discovery potential of these idealized experiments. As expected, for SI WIMP interactions target complementarity has a modest effect and can at most reduce by a factor of 2 the impact of the neutrino background on the discovery potential.

Fig.~\ref{fig:complementarity_SDp} (left panel) is similar to Fig.~\ref{fig:complementarity_SI} (left panel) but for the SD interaction on the proton. From Fig.~\ref{fig:complementarity} (right) we see that fluorine targets have a large advantage in sensitivity for SD proton interactions. 
One can see this in the large improvement in discovery potential when adding $\rm{C_3F_8}$ to the compliment of targets. One might wonder if it is worth combining $\rm{C_3F_8}$ with other experiments due to its advantage for this interaction. To consider this, we have added a $\rm{C_3F_8}$-only discovery and sensitivity limit. One can see that even though one obtains the best background-free sensitivity with $\rm{C_3F_8}$, the best discovery limit is obtained by adding other targets which helps at reducing the effect of the neutrino background by about an order of magnitude (see Fig.~\ref{fig:complementarity_SDp} right panel) even while the background-free sensitivity is reduced. It is also worth recalling that the curves that consider an experiment using $\rm{C_3F_8}$ with recoil energy sensitivity have very similar neutrino background implications than CF$_4$-based experiments such as MIMAC and DMTPC.

Finally Fig.~\ref{fig:complementarity_SDn} shows the results of complimentarity for the SD neutron case. For this interaction, the improvement in discovery potential is intermediate between the SI and SD proton case. 

While we have considered only models in which a WIMP has one of these three types of nuclear interactions, a true WIMP may have non-zero cross sections in all three of these channels. If a potential signal is discovered near the neutrino-induced discovery limit, a more detailed scan of the parameter space would be required to disentangle the possible WIMP and neutrino signals. 

\subsection{Effect on the dynamics of the discovery limit}

In the previous section, we focused on the effect of target complementarity on the discovery potential of upcoming experiments for a fixed exposure. In the following, we will describe the effect of target complementarity on the dynamics of the discovery limit when increasing the exposure around the saturation regime.\\
\indent Figure \ref{fig:evolution} shows the evolution of the discovery limit for Xe (red~$+$), Xe+Ge (blue~$\times$) and Xe+Ge+Si (green~$*$) based experiments considering both SI (left panel) and SD (right panel) interactions for a $6~\rm{GeV/c^2}$ WIMP mass. We can see that in the case of a SI interaction, combining Xe, Ge and Si does not allow one to greatly improve on its discovery potential. The green points (Xe+Ge+Si) still clearly show a saturation regime in the discovery potential. However, considering a SD interaction on the proton, adding successively Ge and Si allows one to consequently improve the evolution of the discovery potential with exposure and keep it close to the $1/\sqrt{MT}$ best-case scenario. By comparing the Xe-only and Xe+Ge+Si points, we can see that target complementarity allows one to reach a discovery limit of about $2\times 10^{-38}~\rm{cm^2}$ with an exposure which is 50 times lower. This clearly highlights the interest of combining data from different experiments to bypass the saturation regime induced by the neutrino background.

\section{Optimization}
\label{sec:optimization}

\begin{figure}
\begin{center}
\includegraphics[width=0.95\columnwidth]{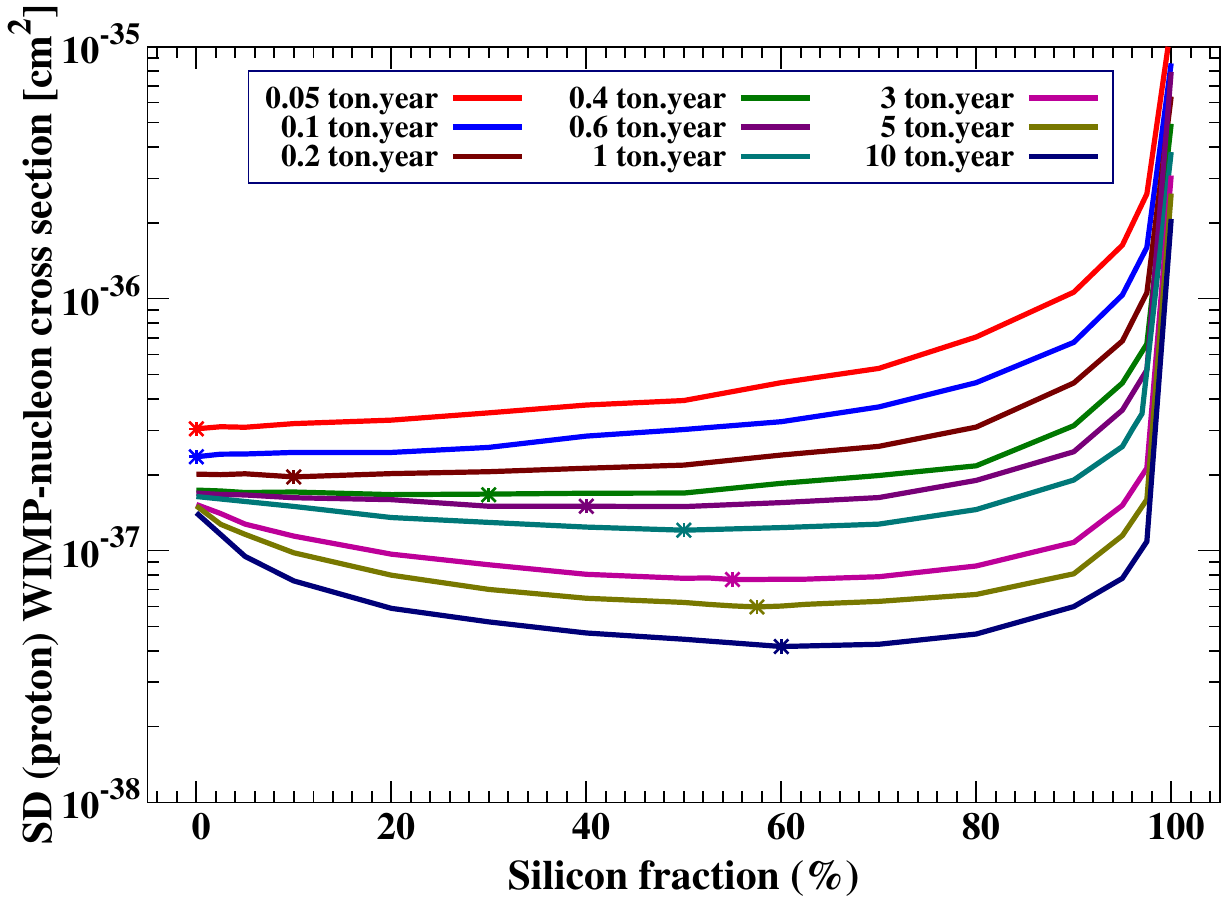}
\end{center}
\caption{Evolution of the discovery limit of a combination of Ge and Si based experiments as a function of the considered Si fraction for several different exposures considering a WIMP mass of $6~\rm{GeV/c^2}$. We considered both the SI and the SD interaction on the proton. For the SI interaction there is no optimal Si fraction because there is no visible effect from target complementarity. However, for the SD interaction on the proton, the discovery limit at $6~\rm{GeV/c^2}$ has a minimum value for a certain Si fraction which depends on the considered exposure.}
\label{fig:optimization}
\end{figure}

The previous results have shown combinations of several single-nucleus based experiments with a fixed relative exposure. In this section, we will address the optimization of several target materials when designing a future experiment that would maximize its discovery potential. We will consider the relative exposures of Si and Ge crystals in this example since at least one experiment (SuperCDMS) can use both of these targets in their future payloads. 

As suggested by our previous study of the target complementarity, in the context of a SI interaction the discovery limit is only slightly affected by changing the Si fraction of the total exposure. We thus concentrate on the SD proton interaction. Figure~\ref{fig:optimization} shows the discovery limit for a combination of Ge and Si-based experiments as a function of the considered Si fraction in the total exposure at a fixed WIMP mass of 6 GeV/c$^2$. The considered thresholds for the two experiments target nuclei are taken from Table~\ref{tab:Exposures}. We can see that for exposures which are low enough to be in the systematics-dominated regime (below 0.2 ton-year), adding Si in the experiment does not improve the discovery potential because the reconstructed cross section for Si is about two orders of magnitude higher than the one for Ge (see Fig.\ref{fig:complementarity}). However, when the exposure is high enough to be in the discovery potential saturation regime, the complementarity of the two targets leads to an improvement of the combined discovery potential. The optimum Si fraction that maximizes the effect of the complementarity of Ge and Si corresponds to the minimum of these curves which has been marked by a ``$*$''. We can therefore see that as the exposure increases, the Si fraction that maximizes the discovery potential has to be higher. For a 0.4~ton-year experiment it is around 30\% but for a 10~ton-year experiment it is around 60\%. Fig.~\ref{fig:optimization} suggests that for very high exposures, it is worth tuning the relative exposures of Ge and Si detectors in order to maximize the discovery potential in light of the neutrino background.

\section{Conclusion}

Since dark matter experiments will soon begin be sensitive to coherent neutrino scattering from solar neutrinos, the search for discrimination methods to disentangle WIMP from neutrino events is a necessity. For both spin-independent and spin-dependent targets, we have examined the effect of target complementarity on the discovery potential of single nucleus and multi-target based experiments. In general, we find that combining different targets such as Xe and Ge will allow for an improvement in the discovery potential. For spin-independent interactions, we show that the similarity between the scaling of the response of WIMPs and neutrinos with different targets ultimately limits the gain in discovery potential achievable. 

However, this situation is different for spin-dependent interactions. In this case, the differences in the cross sections of WIMPs and neutrinos with different targets allows one to bypass the saturation of the discovery potential and allow the search for dark matter to be pursued to lower cross sections. The improvement is most dramatic for spin-dependent WIMP-proton interactions. In particular combining C$_3$F$_8$ with other targets results in a significant decrease of the impact of the neutrino background on the discovery potential.

We have also shown that the effect of target complementarity can be used to optimize the different element fractions for experiments which have more than one target material. In particular for the SD interaction on the proton, there is an optimum Si fraction for a Ge$+$Si experiment which maximizes the discovery potential. We have discussed how this fraction depends on the exposure and increases with the exposure.

In summary, the results that we have presented in this paper provide valuable input for future direct dark matter searches. We have provided the first suggestion for expanding the reach of WIMP dark matter searches in the regime where the neutrino backgrounds are substantial. The methods we have discussed rely only on measuring energy depositions, and can ultimately be complementary to methods that reduce the neutrino background that rely on different experimental signatures such as annual modulation or the direction of the recoiling nucleus. \\

{\bf Acknowledgments}-- LES was supported by NSF grant PHY-1417457. FR, JB and EFF were supported by NSF grant No. NSF-0847342. The authors would like to thank Laurent Derome and Frederic Mayet for useful discussions on an earlier version of this work.

\end{document}